\documentclass[useAMS, usenatbib]{mn2e} 
\usepackage{aas_macros}
\usepackage{graphicx}
\usepackage{epsfig}
\usepackage{subfig}
\usepackage{amsmath}
\usepackage{mathtools}
\usepackage{pdflscape}
\usepackage{longtable}
\usepackage{multicol}
\usepackage{placeins}
\usepackage{pdfpages}
\title{New Pre-main-Sequence Stars in the Upper Scorpius Subgroup of Sco-Cen}
\author[A.C. Rizzuto et al.]{\parbox{\textwidth}{A.C. Rizzuto$^{1,2}$, M.J. Ireland$^{3}$, A.L. Kraus$^1$}
\\
\\
\vspace{0.1cm}
\parbox{\textwidth}{$^{1}$Department of Astronomy, University of Texas, 2515 Speedway, Stop C1400,  Austin, Texas 78712, USA\\
$^{2}$Department of Physics and Astronomy, Macquarie University, Sydney NSW, 2109, Australia\\
$^{3}$Research School of Astronomy \& Astrophysics, Australian National University, Canberra, ACT 2611, Australia\\
}}

\newcommand{\Msun}{M$_{\odot}$~}
\newcommand{\Msunc}{M$_{\odot}$}

\begin{document}

\pagerange{\pageref{firstpage}--\pageref{lastpage}} \pubyear{2013}

\maketitle

\begin{abstract}
We present 237 new spectroscopically confirmed pre-main-sequence K and M-type stars in the young Upper Scorpius subgroup of the Sco-Cen association, the nearest region of recent massive star formation. Using the Wide-Field Spectrograph at the Australian National University 2.3\,m telescope at Siding Spring, we observed 397 kinematically and photometrically selected candidate members of Upper Scorpius, and identified new members  by the presence of Lithium absorption. The HR-diagram of the new members shows a spread of ages, ranging from $\sim$3-20\,Myr, which broadly agrees with the current age estimates of $\sim$5-10\,Myr. We find a significant range of Li 6708 equivalent widths among the members, and a minor dependence of HR-diagram position on the measured equivalent width of the Li 6708\,\AA~ line, with members that appear younger having more Lithium. This could indicate the presence of either populations of different age, or a spread of ages in Upper Scorpius. We also use Wide-Field Infrared Survey Explorer data to infer circumstellar disk presence in 25 of the members on the basis of infrared excesses, including two candidate transition disks. We find that 11.2$\pm$3.4\% of the M0-M2 spectral type (0.4-0.8\,\Msunc) Upper Sco stars display an excess that indicates the presence of a gaseous disk.
\end{abstract}


\section{Introduction}
\label{intro}

The Scorpius-Centaurus-Lupus-Crux Association (Sco OB2, Sco-Cen) is the nearest location to the Sun with recent high-mass star formation \citep{zeeuw99}. Young OB associations, such as Sco-Cen, provide an incredible laboratory in the form of a primordial group of stars directly after formation, which can be exploited in the study of the output of star formation including searches for young exoplanets. The obvious prerequisite for such study is a level of completeness in the identification of association members that is currently not yet attained in Sco-Cen in any mass regime, other than the most massive B-type stars. Sco-Cen contains approximately 150 B-type stars \citep{myfirstpaper} which have been typically split into three subgroups: Upper Scorpius, Upper-Centaurus-Lupus (UCL) and Lower-Centaurus-Crux (LCC) with only the B, A and F-type membership of Sco-Cen being considered relatively complete, with some 800 members. Even in this high-mass regime, there is expected to be a $\sim$30\% contamination by interlopers in the kinematic membership selections, mainly due to the lack of precision radial velocity measurements for these objects \citep{myfirstpaper}.  Additionally, in light of the upcoming high-precision GAIA proper motions and parallaxes, a well characterised spectroscopically confirmed Sco-Cen membership will be instrumental in illuminating the substructure of the association.

Unfortunately, Sco-Cen is poorly characterised for its proximity, the reason for which is the enormous area of sky the association inhabits at low Galactic latitudes ($\sim80^\circ\times25^\circ$ or $\sim150\times50$\,pc).
IMF extrapolation from the high-mass members implies, with any choice of IMF law, that Sco-Cen is expected to have $\sim 10^4$ PMS G, K and M-type members, most of which are, as yet,  undiscovered. This implies that the vast majority of PMS ($<$20\,Myr) stars in the solar neighbourhood are in Sco-Cen \citep{preibisch02}, making Sco-Cen an ideal place to search for young, massive planetary companions. Although some work has been done in illuminating the lower-mass population of Sco-Cen (see \citet{preibisch08}), the late-type membership of Sco-Cen cannot be considered complete in any spectral-type or colour range. A more complete picture of the late-type membership of Sco-Cen is the primary requirement for determining the age spread, structure, and star formation history of the association, for illuminating the properties of star formation, and for embarking on further searches for young exoplanets to better define  their population statistics.

The age of the Sco-Cen subgroups has been contentious. Upper Scorpius has long been considered to be $\sim$5\,Myr old, however recent work has shown that it may be as old as 11\,Myr \citep{geus92, pecaut12}. Similarly, B, A and F-type UCL and LCC members have main-sequence turn off/on ages of $\sim16-18$\,Myr, while studies of the incomplete sample of lithium-rich G, K and M-type members show a variety of mass-dependent age estimates. The HR-diagram age for the known K-type stars in UCL and LCC is $\sim$12\,Myr, the few known M-type stars indicate a significantly younger age of $\sim$4\,Myr, most likely due to a bias produced by a magnitude limited sample, and the G-type members have an age of $\sim$17\,Myr, which is consistent with the more massive stars \citep{preibisch08,song12}. There is also a positional trend in the age of the PMS stars of the older subgroups, with stars closer to the Galactic Plane appearing significantly younger than objects further north. This is almost certainly the result of as yet undiscovered and un-clarified substructure within the older subgroups, which may have a very complex star-formation history.

The above is clear motivation for the identification of the full population of the Sco-Cen association, a task that will require significant observational and computational effort to complete. In this paper, we describe a new search for PMS members of the Upper Scorpius region of the Sco-Cen association. We have used statistical methods to select a sample of likely Upper Scorpius members from all-sky data, and have conducted a spectroscopic survey to determine youth and membership in the Sco-Cen association using the Wide-Field Spectrograph instrument at the Australian National University 2.3\,m telescope. 

\section{Selection of Candidate Members}
We have selected candidate Upper Scorpius members using kinematic and photometric data from UCAC4, 2MASS, USNO-B and APASS \citep{ucac4,2mass,usnob,apass}. A purely kinematic selection of the low-mass members of Sco-Cen is not sufficient to assign membership to G,K and M-type stars because the quality of the astrometric data available would produce an interloper contamination much higher than would be acceptable for future studies using Sco-Cen as an age benchmark. In order to clearly separate young Upper Scorpius members from field stars, spectroscopic follow-up is needed to identify stellar youth indicators. We employed two separate selection methods to prioritise targets based on kinematic and photometric data. 

The first selection used was based on the Bayesian Sco-Cen membership selection of \citet{myfirstpaper}, which uses kinematic and spatial information to assign membership probabilities. We further developed this method to apply to K and M-type stars, in order to properly treat the absence of a parallax measurement. We took the proper-motions from the UCAC4 catalog \citep{ucac4} and photometry from 2MASS and APASS \citep{2mass,apass},  and used the photometry and a premain-sequence isochrone \citep{siess00} to estimate each candidate member's distance. We then treated the proper-motion and estimated distance together to calculate the membership  probability. This selection was magnitude limited, and covered all stars in the UCAC4 catalog with 10$<$V$<$16, and comprised of $\sim$2000 candidate members with membership probability greater than 2\%. For a more complete explanation of the Bayesian selection, including information from \citep{myfirstpaper} and the changes adopted for use with the  K and M-type star data see Appendix \ref{bayesapp}. 

\begin{figure}
\includegraphics[width=0.45\textwidth]{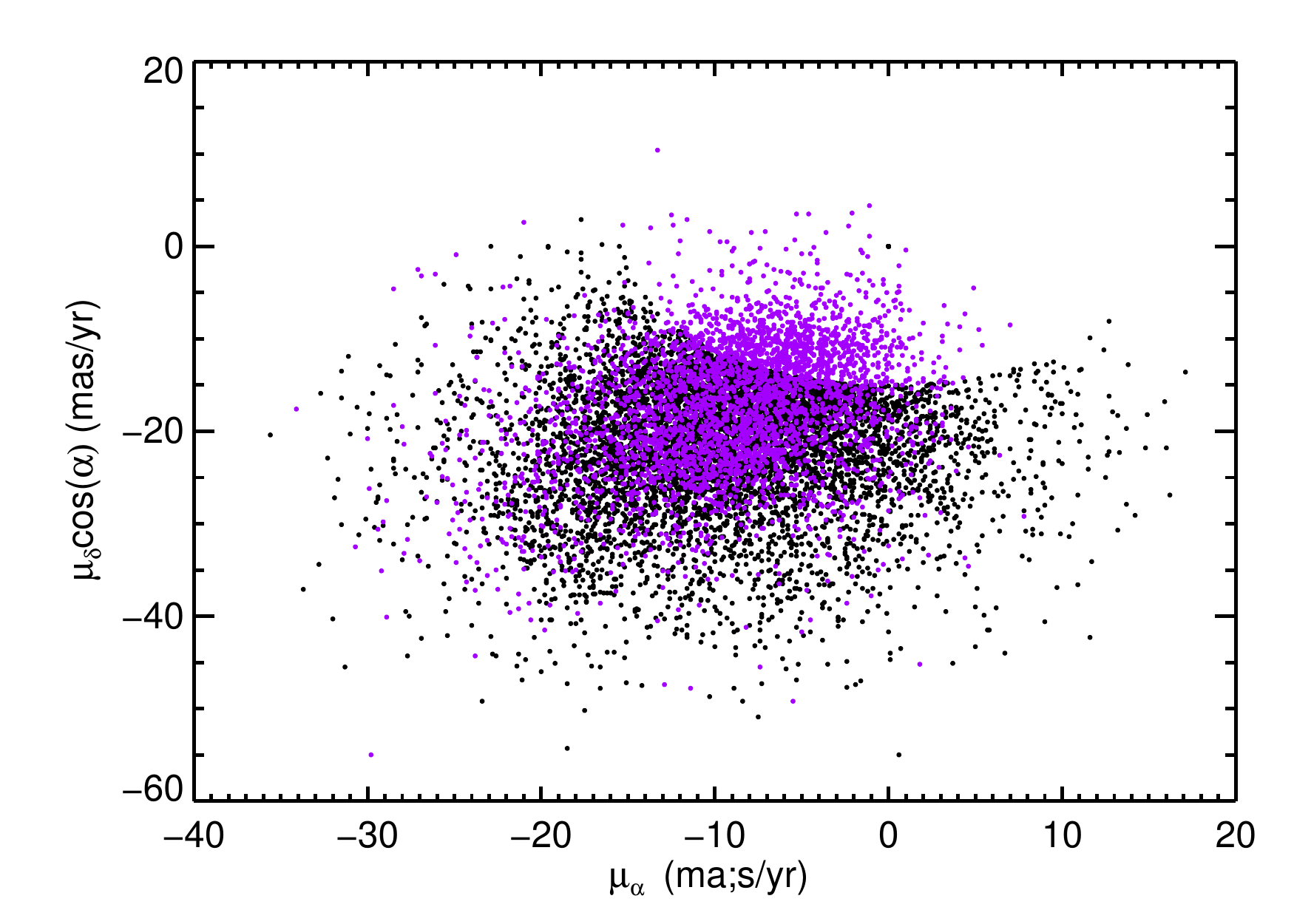}
\caption{The proper motions of the candidate Upper Scorpius members selected by both the \citet{kraushillenbrand_comaber} selection method (black points) and the Bayesian method (purple circles).}
\label{selection_pm}
\end{figure}

The second selection was based on the selection used for the Coma-Ber cluster in the study of \citet{kraushillenbrand_comaber}, and was designed to select targets for the Upper Scorpius field of the Kepler K2 campaign. Targets which were both placed above the main-sequence based on photometric distance estimates, and had proper-motions consistent with Upper Scorpius membership were deemed to be potential members and included in the observing sample. This selection spanned F to late M-type stars, with targets falling on Kepler silicon prioritized for spectroscopic follow-up. This selection is considerably more conservative than the Bayesian selection, and includes a much larger number of candidates. Where the two selections overlap, we have $>$90\% of the Bayesian selected stars included in the sample. Our final, combined sample was then drawn from both of the above selection; we include a candidate in the final target list if it was identified by either method. Figure \ref{selection_pm} displays the proper motions of the selected stars from both samples. 

In light of the currently ongoing Galactic Archeology Survey, using the HERMES spectrograph on the Anglo-Ausralian Telescope \citep{zuckerhermes2012}, which will obtain high-resolution optical spectra in the coming years for all stars in the Sco-Cen region of sky, down to V=14, we have decided to primarily observe targets in our sample fainter than this limit.  While our selection methods identified candidate Upper Scorpius stars across the entire subgroups $(342^\circ<l<360^\circ, 10^\circ<b<30^\circ)$ in our observations we strongly favored candidate members which fell upon the Kepler K2 field 2 detector regions, which covers the majority of the centre of Upper-Scorpius with rectangular windows. As such, the spatial distribution of this sample will not reflect the true substructure of Upper Scorpius. We observed all the targets in our K2 sample with Kepler interpolated V magnitudes of  $(\sim13.5<V_{jk}<15)$, as well as some further brighter targets. In total, we obtained optical spectra for 397 candidate Upper Scorpius K and M-type stars. The full list of observed candidate targets, including both those stars determined to be members and non-members, can be found in Table \ref{obstable_lowmass}, along with proper motions, computed Bayesian membership probability, integration time, and SNR in the continuum near H-$\alpha$.

\begin{table*}
\caption{Summary of WiFeS observations of candidate Upper-Scorpius
members; the V magnitude provided is either taken from APASS, where
available, or interpolated from J and K according to the Kepler K2
instructions. The full table is provided in the online material.}
\label{obstable_lowmass}
\begin{tabular}{cccccccccccc}
\hline
R.A. & Decl. & & V & K & $\mu_\alpha$ &$\mu_\delta$  &  &  & T & & \\
(J2000.0) & (J2000.0) & MJD & (mag) & (mag) &(mas) & (mas) & Source  & P$_{\mathrm{mem}}$  & (sec) & SNR & M? \\
\hline
15 39 06.96 &  -26 46 32.1 & 56462 & 12.5 & 8.7 & -35.3 & -41.7 & a & 31 & 90& 131 & Y\\
15 37 42.74 &  -25 26 15.8 & 56462 & 13.5 & 9.7 & -14.6 & -26.7 & a & 85 & 90 & 80 & \\
15 35 32.30 &  -25 37 14.1 & 56462 & 11.7 & 8.4 & -9.0 & -22.9 & a & 69 & 90 & 116 & \\
15 41 31.21 &  -25 20 36.3 & 56462 & 10.0 & 7.2 & -16.9 & -28.7 & a & 86 & 90 & 151 & Y\\
\hline
\end{tabular}
\end{table*}

\section{Spectroscopy with WiFeS} 

The Wide-Field Spectrograph (WIFES) instrument on the Australian National University 2.3m telescope is an integral field, or imaging, spectrograph, which provides a spectrum for a number of spatial pixels across the field of view using an image slicing configuration. The field of view of the instrument is 38$\times$25 arcseconds, and is made up of 25 slitlets which are each one arc second in width, and 38 arcseconds in length. The slitlets feed two 4096$\times$4096 pixel detectors, one for the blue part of the spectrum and the other for the red, providing a total wavelength coverage of 330 - 900\,$\mu$m, which is dependent on the specific gratings used for the spectroscopy. Each 15 micron pixel corresponds to 1$\times$0.5 arcseconds on sky.  

There are a number of gratings offered to observers for use with WiFeS. For identification of Upper Scorpius members, we required intermediate-resolution spectra of our candidate members, with a minimum resolution of $\sim$3000 at the Li 6708\,\AA~ line, and so selected the R7000 grating for the red arm and the B3000 grating for the blue arm, which was used solely for spectral-typing. This provided $\lambda / \Delta\lambda\sim7000$ spectra covering the lithium 6708\AA~ and H-$\alpha$ spectroscopic youth indicators. A dichroic, which splits the red and blue light onto the two arms of the detector, can be position either at 4800\,\AA~or 5600\,\AA. For the first three successful observing nights we use the dichroic at 4800\,\AA~ which produced a single joined spectrum from 3600 to 7000\,AA. For the remaining 7 nights, we position the dichroic at 5600\,\AA~which produces two separate spectra, with the the blue arm covering 3600 to 4800\,\AA~ and the red arm covering 5300 to 7000\,\AA. This change was made to accommodate poor weather backup programs being simultaneously carried out, which will be the subject of future publications. To properly identify members, we required a $3\sigma$-detection of a 0.1\AA~ equivalent width Li line, which corresponds to a signal-to-noise ratio of at least 30 per pixel. In order to achieve this, we took exposures of 5 minutes for R=13 stars (approximately type M3 in Upper Scorpius), and binned by 2 pixels in the y-axis, to create 1$\times$1" spatial pixels and reduce overheads. With overheads we were able to observe 10 targets an hour in bright time, or $\sim$80-90 targets per completely clear night.

In total we obtained 18 nights of time using WiFeS, split over 2013 and 2014, however the majority of the 2013 nights were unusable due to weather. Our first two observing runs, in June 2013, and April 2014 yielded one half-night of observations each, and our final observing run yielded seven partially clear nights.  During our first two nights, we positioned the dichroic at 5500\,\AA, and during the June 2014 observing run, positioned the dichroic at 4600\,\AA, which provides more of the red arm, because this mode was deemed better for obtaining radial velocities of B, A and F-type Sco-Cen stars, which we observed as backup targets during poor weather, and will be the subject of a future publication.

\section{Data Reduction}
The raw WiFeS data was initially reduced with a pre-existing Python data reduction software  package called the ``WiFeS PyPeline'', which was provided to WiFeS observers. The purpose of the software is to transform the CCD image, which consists of a linear spectrum for each spatial pixel of the WiFeS field of view, into a data cube. This involves bias subtraction, flat-fielding, bad pixel and cosmic ray removal, sky subtraction, wavelength calibration, flux calibration, reformatting into the cube structure, and interpolation across each pixel to produce a single wavelength scale for the entire image. On each night, we observed at least one flux standard from \citet{bessellflux99}, which are included in the data reduction pipeline as flux calibrator objects. Once this process is complete, the user is left with a single cube for each object observed, with dimensions 25''$\times$38''$\times$3650 wavelength units. For the grating resolutions and angles used in our observations, we obtained spectral coverage from $3200-5500$\,\AA~in increments of 1.3\,\AA~in the blue arm, and $5400-7000$\,\AA~ in increments of 0.78\,\AA~in the red arm. 

\begin{figure}
\centering
\subfloat[\label{spect_example}]{\includegraphics[width=0.45\textwidth]{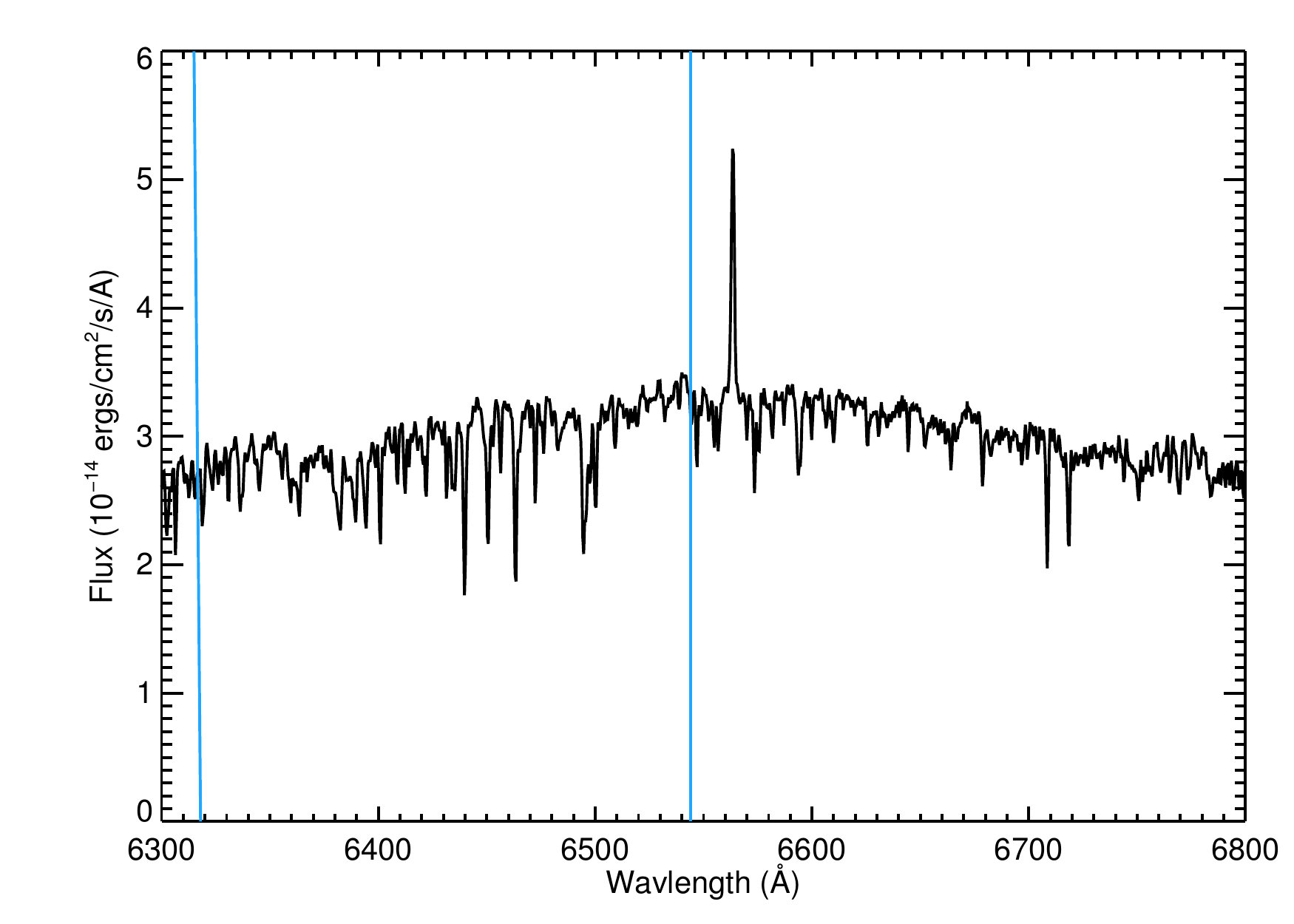}}\\
\subfloat[\label{subim_example}]{\includegraphics[width=0.45\textwidth]{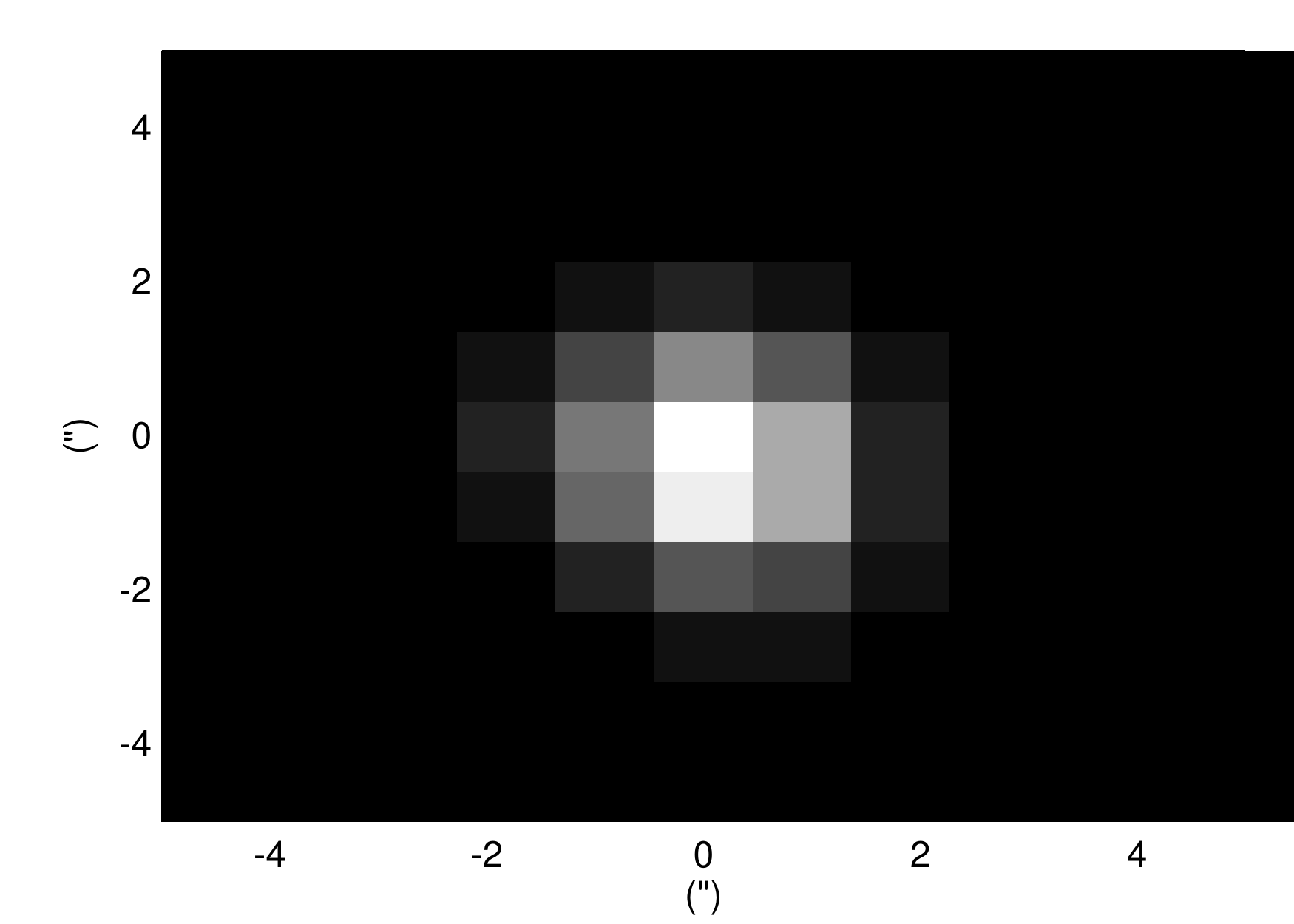}}
\caption[WiFeS spectrum and spatial image of 1RXS J153910.3-264633, a high priority target in our observations sample]{(a) Example spectrum for object 1RXS J153910.3-264633, a high priority target in our observation sample, which shows signs of youth such as H-$\alpha$ emission and Li 6708\AA~absorption. The region of the continuum used for the initial PSF fitting is bounded by blue lines. (b) Spatial image created for  1RXS J153910.3-264633 by adding the images at each wavelength of the PSF fitting region of the continuum.}
\label{red_examle}
\end{figure}

Following the standard WiFeS reduction procedure, we continued with a further custom reduction, the aim of which was to measure the centroid position of the target object in each wavelength, such that the presence of  H-$\alpha$ emitting low-mass stellar companions, outflows, and H-$\alpha$-bright planetary mass companion could be detected by the measurement of a wavelength-dependent centroid shift.  This consisted of determining a best fit point spread function (PSF) model for the spatial image in a clean section of the spectrum, and then measuring the centroid shift of this PSF at each wavelength along the spectrum. An additional benefit of this is a more accurate sky subtraction, and an integrated spectrum of each object, which can be used to measure equivalent widths of key spectral lines. The results of the centroid measurements and any detected companions will be reported in a further publication. 

We first cut out a 10'' by 10'' wide window (10$\times$10 pixels), centered on the target. The vast majority of the stellar flux is contained within the central 3'' by 3'' region of the windowed image, and so the adopted width of 10'' allows a clear region of background around the target; Figure \ref{subim_example} provides an illustration of the data. We then fit a Moffat point spread function \citep{racine96} to a region of the spectral continuum which does not include any spectral features, but is close to the H-$\alpha$ line. This region consisted of 400 spectral units, spanning $6368-6544$\,\AA. Figure \ref{spect_example} displays the spectral region used for the initial PSF fit, as well as the H-$\alpha$ and Li 6708\,\AA~lines for one target in our sample, 1RXS J153910.3-264633, which shows strong indications of youth.

The particular model that we fit to the spatial image is given by;

\begin{equation}
\mathrm{PSF} = \mathrm{S} + \mathrm{F} \frac{(2^{\frac{1}{\beta}}-1)(\beta-1)}{\pi w^2(1 + (2^{\frac{1}{\beta}}-1)(\theta/w)^2)^{\beta}},
\label{moffat}
\end{equation}

where S indicates the sky contribution to the flux, $\beta$ is an integer parameter that determines the strength of the wings of the Moffat PSF,  $\theta$ is the distance from the centre of the profile, $w$ is the half width of the Moffat PSF, and F is the stellar flux. Given that we have a two dimensional PSF, and that each dimension has a different Moffat function half width, we require two different values of $w$. We create this two dimensional Moffat profile by scaling $\theta$ appropriately;

\begin{equation}
\theta = w_x^2(x-x_0)^2 + w_y^2(y-y_0)^2,
\label{moff_scale}
\end{equation}

where $w_x$ and $w_y$ are the PSF width parameters in each dimension, $(x,y)$ is the position of a given point on the image, and $(x_0,y_0)$ is the image centroid. Inputting this value of $\theta$ into a Moffat function with width $w=1$ will thus produce the desired asymmetric two dimensional profile.

\begin{figure}
\centering
\includegraphics[width=0.48\textwidth]{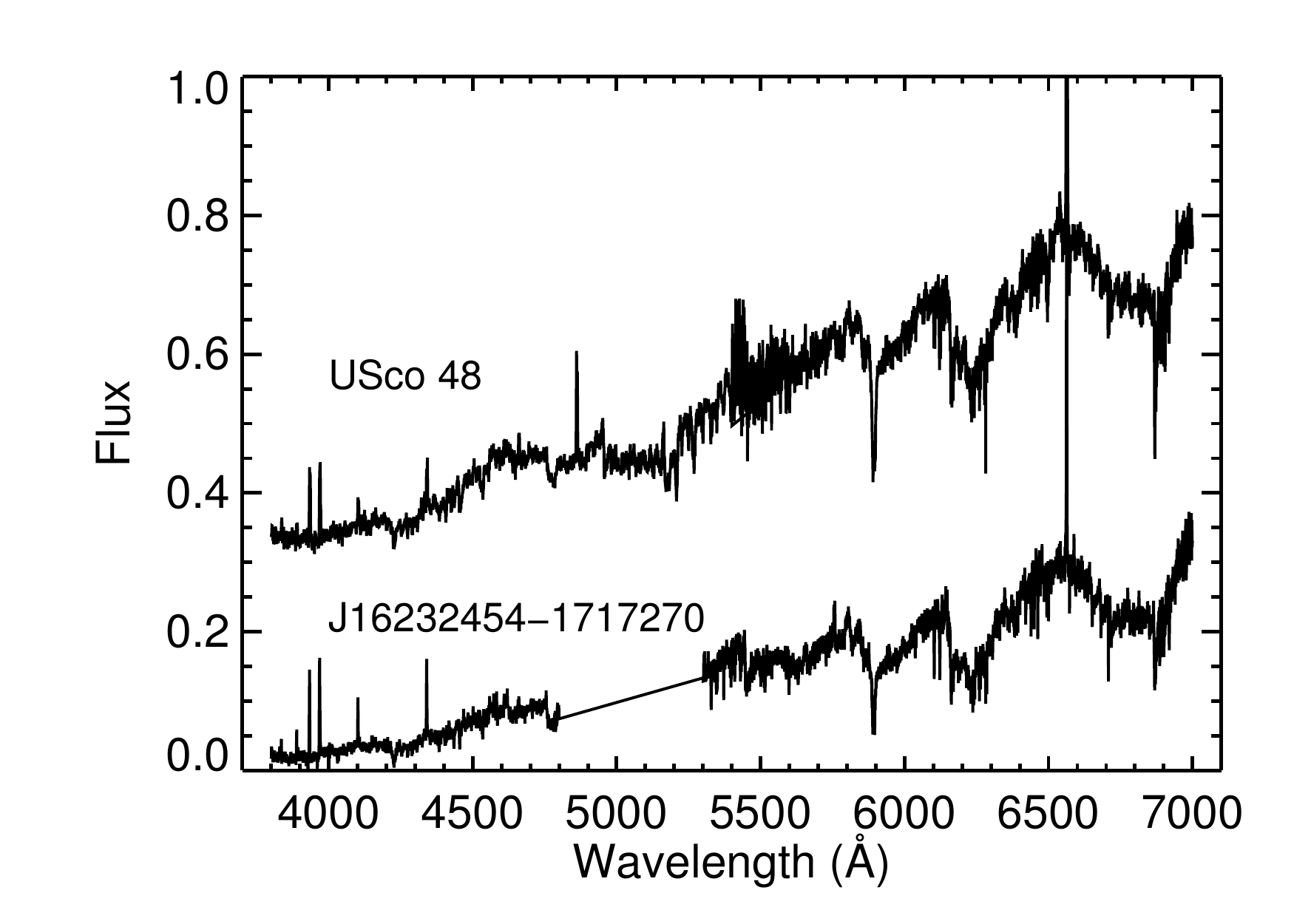}
\caption{The full WiFeS integrated spectrum produced by first processing with the WiFeS Pypeline, and then our spectro-astrometric analysis for the stars USco 48, a known member of the Upper Scorpius subgroup, and 2MASS J16232454-1717270, high probability candidate, and a new member in identified in our survey. The USco 48 spectrum is an example of the data from the 4800\,\AA~dichroic setup, and the 2MASS J16232454-1717270 spectrum an example of the 5600\,\AA~dichroic setup.}
\label{fullspec_example}
\end{figure}

We found that $\beta=4$, a value which describes most telescope PSFs, yielded the closest fit to our data. We also attempted to fit a Gaussian profile to the spatial images, in the same format as the Moffat profile described in equation \ref{moffat}; however the Gaussian model produced consistently poorer fits to the data than the Moffat model, particularly in the wings of the PSF, with typical values of  $\chi^2_r\sim 4$ for the Gaussian model fit and $\chi^2_r\sim 2$ for the Moffat model. On the basis of the goodness of fit difference, we adopted the Moffat model exclusively in our analysis. For each target observed, we used the continuum spectral region between $6368-6544$\,\AA~to determine the parameters of the Moffat PSF that most closely reproduced the spatial images. We then fixed the half width parameters in each dimension, and fit our PSF model to each individual wavelength element image along the spectrum to determine $S$, $F$ and the centroid position for each wavelength. This process provides two useful characteristics, the first of which is the integrated spectrum $(F)$ of the target (see Figure \ref{fullspec_example}), with the sky component $(S)$ subtracted out. Using the cleaned output spectra, we then computed equivalent widths of both the Li 6708\,\AA~ and H-$\alpha$ lines for each observed star. The second useful characteristic is the centroid position of the star image at each wavelength interval in the spectrum. This can be used to detect accreting stellar and substellar companions by the measurement of a centroid shift in the H-$\alpha$ line image. An analysis of the centroid positions will be presented in a future publication.

\begin{figure*}
\subfloat[\label{newmems_pos}]{\includegraphics[width=0.45\textwidth]{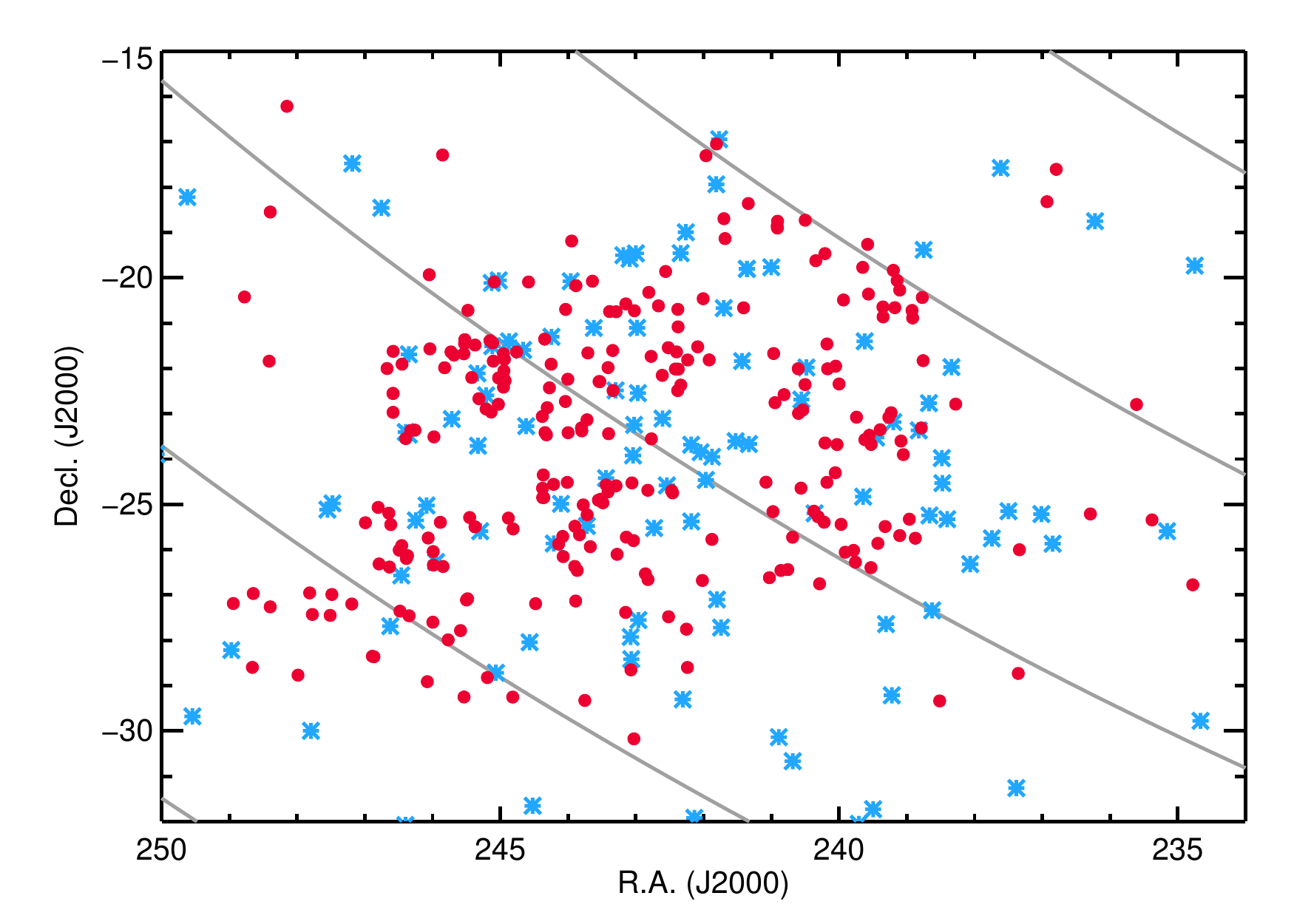}}
\subfloat[\label{pm_all}]{\includegraphics[width=0.45\textwidth]{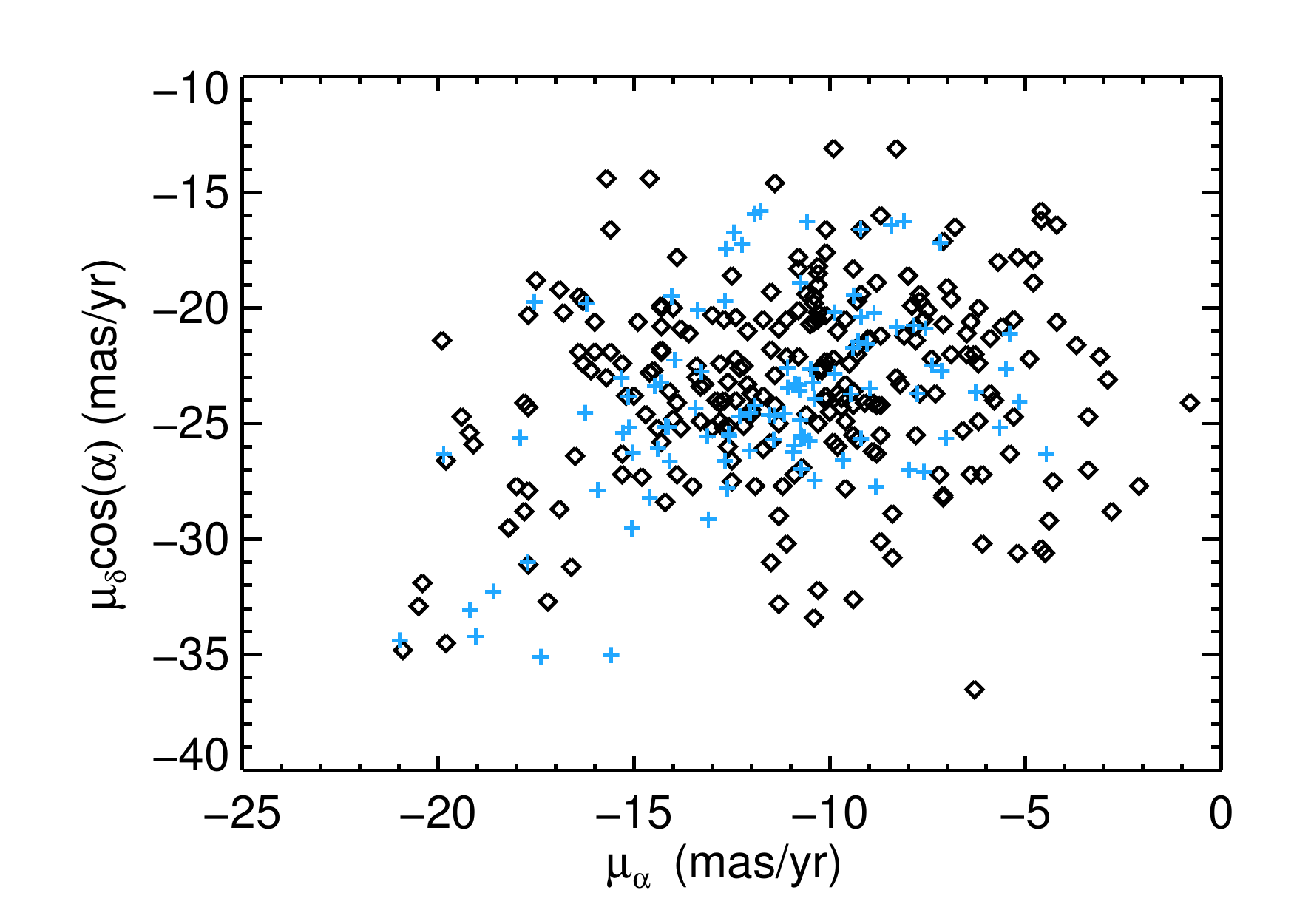}}
\caption{(a) On-sky positions of the new members (red circles), relative to the B, A and F-type members (blue stars) from \citet{myfirstpaper}. Lines of constant Galactic latitude are shown in grey in steps of 10$^\circ$, the centre line is $b=20^\circ$.  Note that the apparent substructure seen in the new members is artificially created because we strongly prioritized the Kepler K2 Field 2 detector regions in our survey. (b) The proper motions of the new members (black) and the Hipparcos Upper Scorpius members from \citet{myfirstpaper} (blue crosses), the typical proper motion uncertainty for our new members is 2-3\,mas/yr. There is one new member off-scale at (-35.3,-41.7)\,mas/yr. Our new members occupy the same region of proper motion space as the established high-mass members.}
\label{newmems_pos}
\end{figure*}

\subsection{Spectral Typing}

We spectral type the reduced spectra created by the centroid-fitting procedure using spectral template libraries as reference.  It is also important to incorporate extinction into the spectral-typing procedure for Upper Scorpius, given the typical values of $0.5<$A$_V<2.0$. If an extinction correction is omitted, spectral typing will produce systematically later spectral types for the members. A combination of two template libraries was chosen for the spectral typing, with spectral types earlier than M0 taken from the \citet{pickles98} spectral template library, and the M-type templates taken from the more recent \citet{bochanski_templates}. 

To carry out the spectral typing, we first computed reduced $\chi^2$ values for each data spectrum on a two-dimensional grid of interpolated template spectra and extinction, with spacing of half a spectral sub-type and 0.1 magnitudes in $E(B-V)$. This was done by first interpolating the template spectra onto the wavelength scale of the data, and then applying the particular amount of extinction according to the \citet{savage_mathis79} extinction law.  We also removed the H-$\alpha$ region in the data spectra, because the prevalence of significantly larger H-$\alpha$ emission in young stars will not be adequately reproduced by the templates. The spectral type - extinction point on the grid with the smallest reduced $\chi^2$ was then used as a starting point for least squared fitting with the IDL fitting package MPFIT. The fitting procedure used the same methodology as the grid calculations, with the addition of interpolation between template spectra to  produce spectral sub-type models for use in the fitting.

We find the limiting factor in  spectral-typing our young Sco-Cen stars to be the fact that the spectral template libraries are built from field stars, and so are not ideal for fitting young, active stars. Hence, while we typically have spectral type fits better than half a spectral sub-type, we report spectral types to the nearest half sub-type, and values of A$_V$ with typical uncertainties of 0.2 magnitudes.

\section{The New Members}

\begin{figure*}
\centering
\subfloat[\label{ewli_all}]{\includegraphics[width=0.45\textwidth]{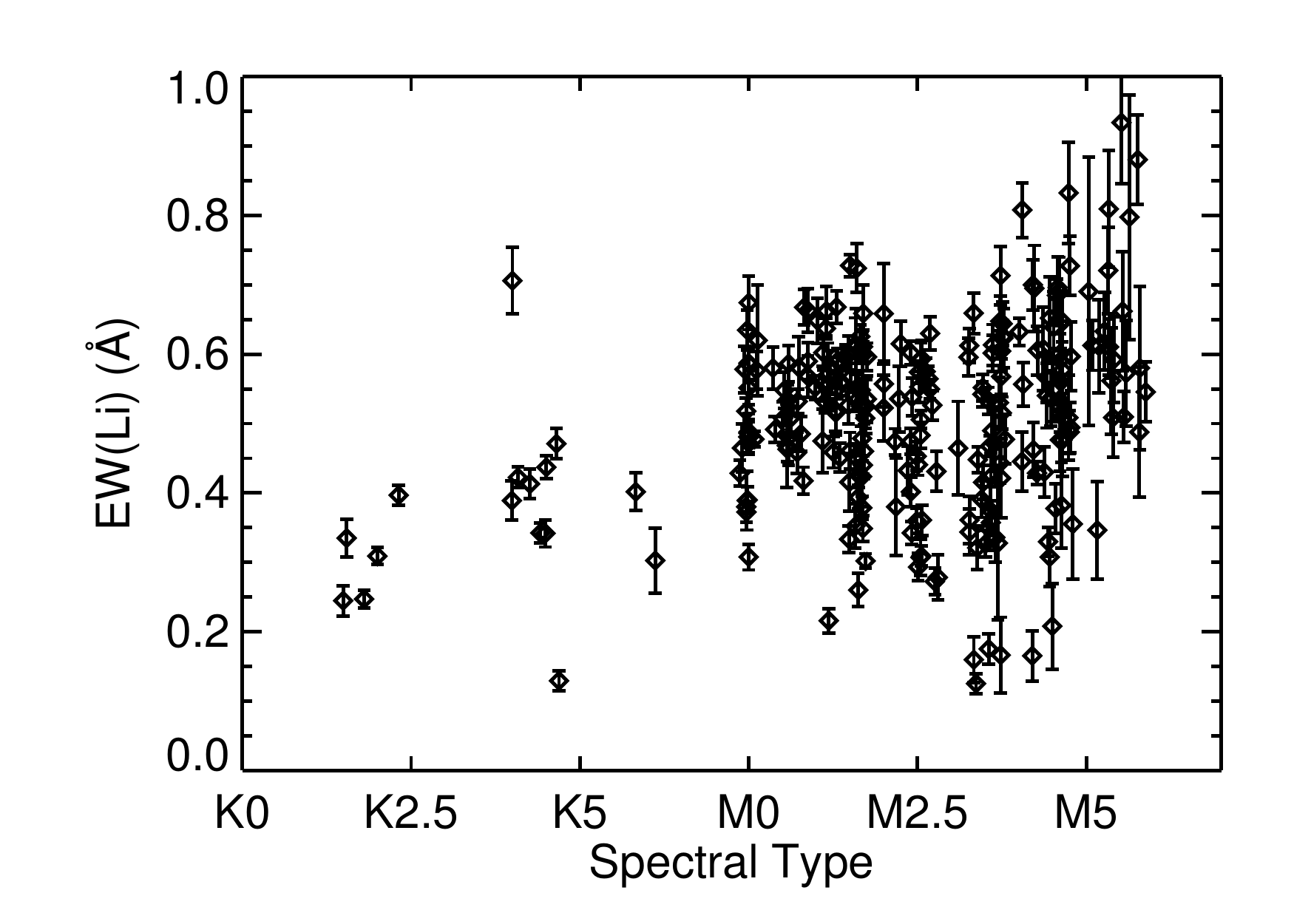}}
\subfloat[\label{ewli_zoom}]{\includegraphics[width=0.45\textwidth]{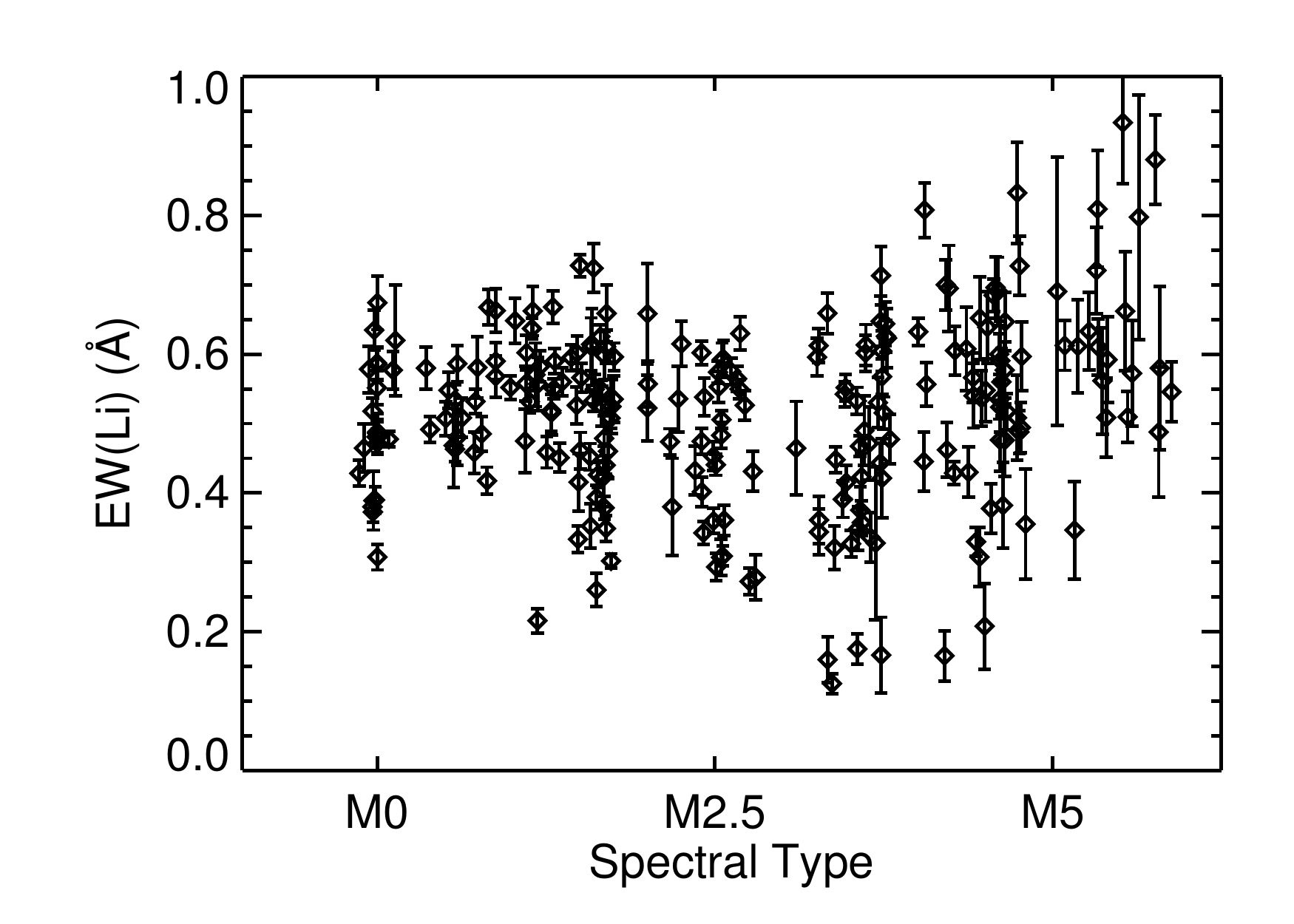}}
\caption{$\mathrm{EW}(\mathrm{Li})$ for the new members. The left panel shows the full K to late M-spread, while the right panel focusses on the M-type range, which shows the largest spread in EW(Li) for a given spectral type.}
\end{figure*}

\begin{table*}
\caption{Properties of the Upper Scorpius members identified in our survey. The first column lists our adopted naming system for these new members and the second column lists the 2MASS designation for each of our targets. We also list the the fitted spectral type and visual extinction,  and the equivalent widths of the Li 6708\AA~and H$\alpha$ lines (EW(Li) and EW(H$\alpha$)). The full table is provided in the online material.}
\label{obs_res_tab}
\begin{tabular}{l c c c c c c c c c}
\hline
& & R.A. & Decl. & EW(Li) & $\sigma_{EW(Li)}$& EW(H$\alpha$) &$\sigma_{\mathrm{EW(H_{\alpha})}}$ & &  A$_{\mathrm{V}}$  \\
Name &2MASS & (J2000.0) & (J2000.0) & (\AA) & (\AA) & (\AA) & (\AA) & SpT & (mag) \\
\hline
RIK-1 & J15390696-2646320 & 15 39 06.96 &  -26 46 32.1 & 0.46 & 0.02 & -1.22 & 0.03 & M0.5 & 0.2\\
RIK-2 & J15413121-2520363 & 15 41 31.21 &  -25 20 36.3 & 0.40 & 0.01 & -2.70 & 0.04 & K2.5 & 0.1\\
RIK-3 & J15422621-2247458 & 15 42 26.21 &  -22 47 46.0 & 0.46 & 0.04 & -3.08 & 0.07 & M1.5 & 0.3\\
RIK-4 & J15450970-2512430 & 15 45 09.71 &  -25 12 43.0 & 0.61 & 0.02 & -2.02 & 0.04 & M1.5 & 0.4\\
\hline
\end{tabular}
\end{table*}

Table \ref{obs_res_tab} lists both the Li 6708\,\AA~ and H$\alpha$ equivalent widths, and the estimated spectral types and extinction for the new Upper Sco members, and figure \ref{newmems_pos} shows the spatial positions of the new members. We have defined a star as an Upper Scorpius member if the measured equivalent width of the Li 6708\,\AA~ line was more than $1-\sigma$ above 0.1\,\AA.  While this Li threshold is low, it is significantly larger than the field Li absorption, and is in general keeping with previous surveys. The use of this threshold is further justified given the effects of episodic accretion on Li depletion in the latest models \citep{baraffe10}. In general, the vast majority of the identified members have Li 6708\,\AA~ equivalent width significantly larger than 0.2\,\AA~and so are bonafide young stars. In total we identify 257 stars as members based on their Li 6708\,\AA~ absorption, 237 of which are new. 

The proper-motions of the new members, which were calculated from various all-sky catalogs, or taken from the UCAC4 catalog are shown in Figure \ref{pm_all}. The members have proper motions that overlap the Upper Scorpius B, A and F-type members proper motions (blue crosses), although a significantly large spread is seen. This is consistent with the average uncertainty of $\sim$2-3\,mas/yr for the K and M-type proper motions. 

Figure \ref{ewli_all} displays the Lithium equivalent widths for the identified members as a function of spectral type. The majority of our members are M-type, and we see a sequence of equivalent width with a peak at spectral type M0, and a systematically smaller equivalent width in the M2-M3 range compared to earlier or later M-type members. This is expected as the mid-M range is modelled to show faster Lithium depletion timescale \citep{dantona94}.  

Interestingly, we also observe a clear spread in the equivalent width of the Lithium 6708\AA~ line. Figure \ref{ewli_zoom} shows the just the M0 to M5 spectral type range. At each spectral type we see a typical spread of $\sim$0.4\,\AA~in Li equivalent width, and a median uncertainty in the equivalent width measurements of $\sim$0.03\,\AA. This implies a $\sim$10-sigma spread in EW(Li) at each spectral type. Wether or not this spread is caused by an age spread in Upper Scorpius is difficult to determine: we have examined the behaviour of EW(Li) as a function of spatial position, both in equatorial and Galactic coordinate frames and found no significant trend. We note that a similar spread of EW(Li) for M-type Upper Scorpius members was observed by \citet{preibisch01}. Given the lack of correlation with spatial position, if the EW(Li) spread is caused by an age spread among the members, then the different age populations are overlapping spatially and may not be resolvable without sub-milliarcsecond parallaxes.

\begin{figure}
\centering
\includegraphics[width=0.47\textwidth]{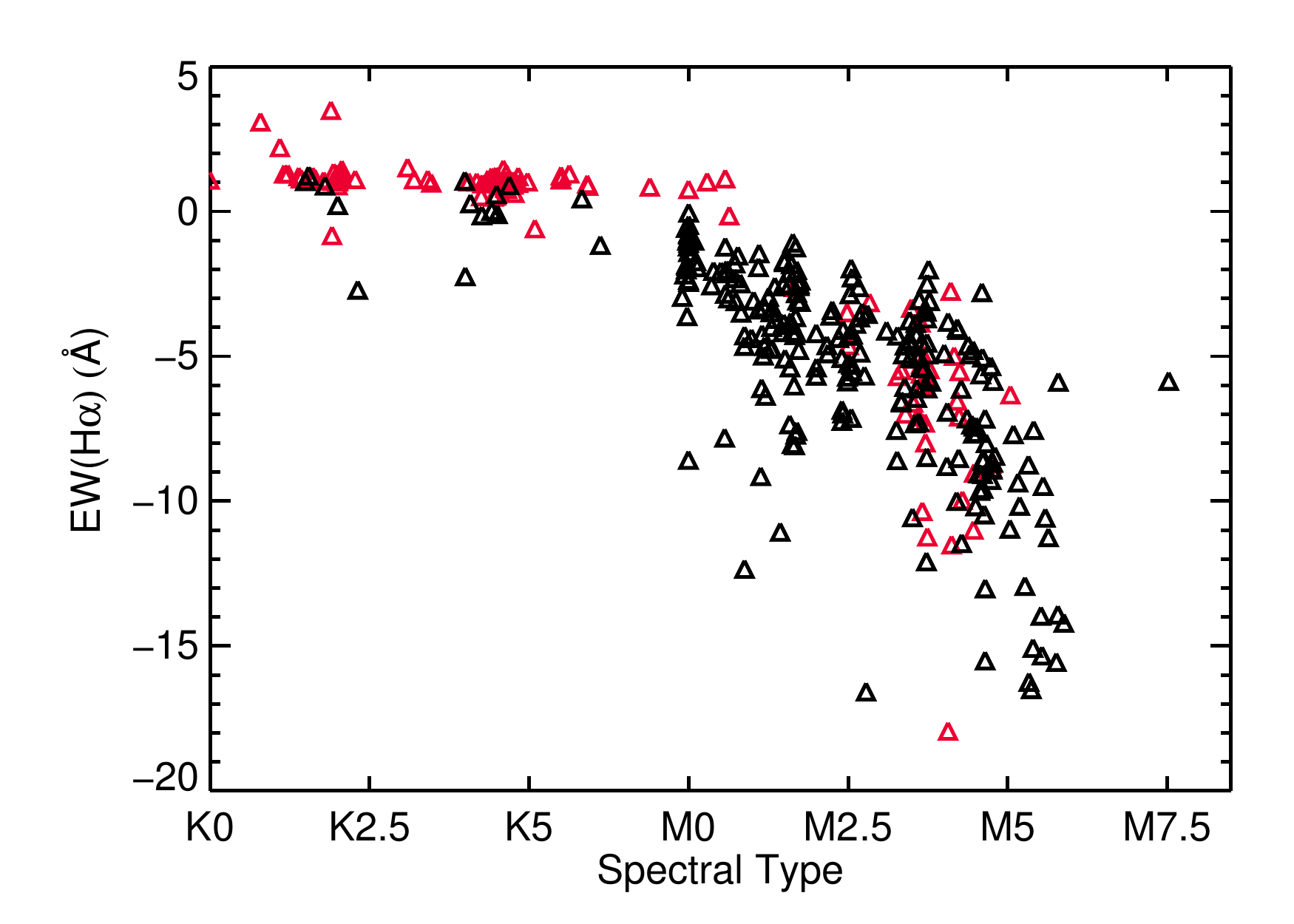}
\caption{$\mathrm{EW}(\mathrm{H-}\alpha)$  for the new members (black) and the non-members (red). The members follow a clear sequence with H-$\alpha$ increasing with spectral-type. In the K spectal types, we see that non-members show H-$\alpha$ absorption which is generally stronger than that seen in the members, some of which show weak emission.}
\label{ewha_all}
\end{figure}

In Figure \ref{ewha_all}, we display the measured H-$\alpha$ equivalent widths for the members. The majority of the PMS members show some level, of H-$\alpha$ emission, with a clear sequence of increasing emission with spectral type. In combination with the presence of Lithium, this is a further indicator of the youth of these objects. Of our 257 members, $\sim$95\% show H-$\alpha$ emission with (1\,\AA$<$EW(H-$\alpha$)$<$10\,\AA), and only 11 of the members do not show emission in H-$\alpha$. All of these 11 members without H-alpha emission are earlier than M0 spectral-type. There are also 35 non-members with H-$\alpha$ emission.  
Given the values of EW(H$\alpha$) for the M-type members we have identified, the majority of them appear to be  weak lined T-Tauri stars and $\sim$10\% are Classical T-Tauri stars (CTTS) with EW(H$\alpha$) $>$ 10\,\AA. This proportion agrees with previous studies of Upper-Scorpius members \citep{walter94,PZ99,preibisch01}, which find a CTTS fraction of between 4 and 10\% for K and  M-type Upper-Scorpius stars.

\section{The Efficiency of the Bayesian Selection Algorithm}
\begin{figure}
\centering
\includegraphics[width=0.45\textwidth]{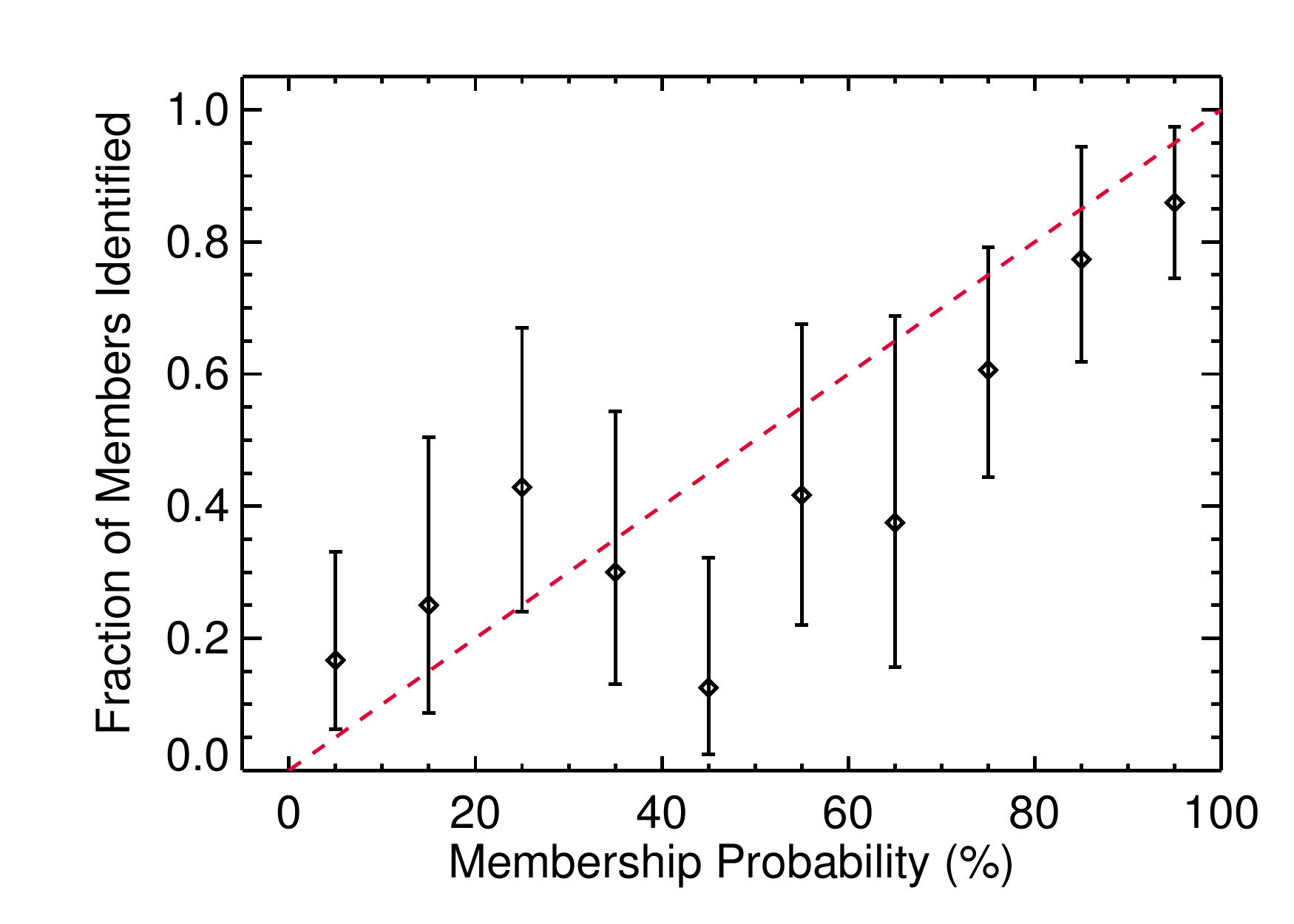}
\caption{Fraction of stars identified as members plotted against membership probability computed with our Bayesian selection algorithm. The red line represents the ideal fraction of detected members. We see a very close agreement between the computed membership probability and the fraction of stars which were confirmed as members.}
\label{bayes_hist}
\end{figure}

The selection methods we have used to create our target list provide a significant improvement of member detection rate when compared to what can be achieved from simple color-magnitude cuts. We see a large overall identification rate of $\sim$65\% for our sample of observed stars. Using the membership probabilities computed for the stars we have observed, we expected that 73$\pm$7\% of the observed stars would be members, which agrees with the observed members fraction of 68\%. We also find that as a function of computed membership probability, the fraction of members identified among the sample behaves as expected. Figure \ref{bayes_hist} displayed the membership fraction as a function of probability. 



Given that our probabilities have been empirically verified to provide a reasonable picture of Upper Scorpius membership, we can derive an estimate for the expected number of M-type members in the subgroup by summation of the probabilities. We find that the total expected number of Upper Scorpius members in the $\sim$0.2 to 1.0\,\Msun range, or late-K to $\sim$M5 spectral type range, is $\sim$2100$\pm$100 members. This agrees with initial mass function estimates which indicate that there are $\sim$1900 members with masses smaller than 0.6\,\Msun in Upper Scorpius \citep{preibisch02}.

\section{The HR-Diagram of the Members}
\begin{figure}
\centering
\includegraphics[width=0.48\textwidth]{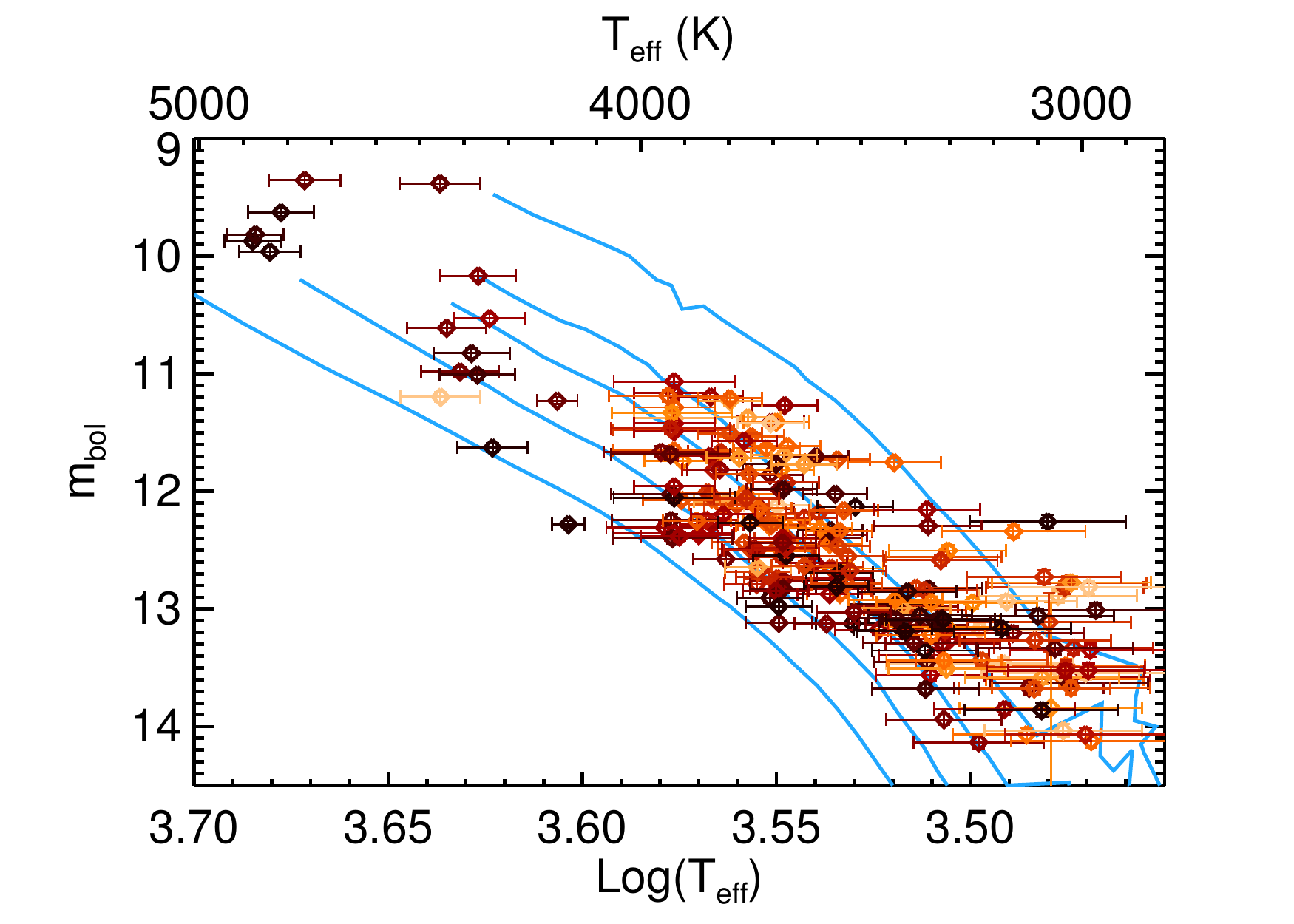}
\caption{HR diagram for the Upper Scorpius members we have identified, with bolometric corrections and effective temperatures taken from the \citet{pecaut13} young star temperature-color scale. The blue lines are the BT-Settl isochrones \citep{btsettl} of ages 1, 3, 5, 10 and 20\,Myr placed at the typical distance to Upper Scorpius of 140\,pc. The color of each point indicates the measured EW(Li) for the star, with darker color indicating a lower EW(Li). The color range spans $0.3<$EW(Li)$<0.7$ linearly, with values outside this range set to the corresponding extreme color. The uncertainties are determined by the accuracy of our spectral typing methods, which is typical half a spectral sub-type.}
\label{cmd_all}
\end{figure}

With the spectral types and extinctions we have determined for the members using the \citet{bochanski_templates} and \citet{pickles98} spectral libraries, we can place them on a HR-diagram in the model parameter space. There is significant variability in synthesized photometry between different models for PMS stars, making comparison in the color-magnitude space difficult. Furthermore, the most reliable magnitudes for M-type stars are the near-IR 2MASS photometry, which show minimal variation in the M-type regime where the PMS is near vertical. Instead, we use the spectral types and the empirical temperature scale and J-band bolometric corrections for 5-30\,Myr stars produced by \citet{pecaut13} and we further correct for extinction using our fitted values of A$_V$ from the spectral typing process, and the \citet{savage_mathis79} extinction law. The resulting HR diagram can be seen in Figure \ref{cmd_all}. We have also superimposed five BT-Settl \citep{btsettl} isochrones of ages 1, 3, 5, 10 and 20\,Myr onto the HR-diagram at the typical Upper Scorpius distance of 140\,pc \citep{myfirstpaper}. These particular models were chosen because they were used by \citet{pecaut13} in the generation of their temperature scale, and so any relative systematic differences between the models and the temperature scale will most likely be minimized.

Upon initial inspection, it appears that for a given temperature range, the Upper Scorpius members inhabit a significant spread of bolometric magnitudes. This is most likely highly dominated by the distance spread of the Upper Scorpius subgroup, which has members at distances between 100 and 200\,pc, corresponding to a spread in bolometric magnitude of $\sim$1.5\,mag between the nearest and furthers reaches of Upper Scorpius. Using the distance distribution of the \citet{myfirstpaper} high-mass membership for Upper Scorpius, we find that the expected spread in bolometric magnitude due to distance which encompasses 68\% of members is approximately $+0.33$ and $-0.54$ magnitudes. Similarly, unresolved multiple systems can bias the sample towards appearing younger by an increase in bolometric magnitude of up to $\sim$0.7\,mags for individual stars.  

In the later spectral types, beyond $\log{T_{\mathrm{eft}}}=3.52$ we also begin to see the effects of the magnitude limit of our survey, which operated primarily in the range $13.5<$V$<15$ and so only the brightest, and hence nearest and potentially youngest late M-type members in our original target list were identified, although significant Li depletion at these temperatures is not expected to occur until ages beyond 50\,Myr. Even with distance spread blurring the PMS in Upper Scorpius, we can see that most of the members appear to be centered around the 5-10\,Myr age range in the earlier M-type members. 

We have also indicated the measured EW(Li) values for the members on the HR-diagram as a color gradient, with darker color indicating a smaller EW(Li). The scale encompasses a range of $0.3<$EW(Li)$<0.7$\,\AA, with values outside this range set to the corresponding extreme color. There is a marginal positional dependence of HR-diagram position with EW(Li): we see that, in particular for the earlier M-type members, the larger values of EW(Li) (light orange) are more clustered around the 3-5\,Myr position, while the smaller values of EW(Li) (dark red) are clustered closer to 5-10\,Myr. This could indicate the presence of a spread of ages, or populations of different age in the Upper Scorpius subgroup.

There is some other evidence of different age populations in the Upper Scorpius subgroup:  The existence of very young B-type stars, such as $\tau$-Sco, and $\omega$-Sco which have well measured temperatures and luminosities that indicate an age of $\sim$2-5\,Myr \citep{simondiaz06} support a young population in Upper Scorpius. 
The B0.5 binary star $\delta$-Sco is also likely to be quite young ($\sim$5\,Myr) \citep{code76}. \citet{pecaut12} place it on the HR diagram at and age of $\sim$10\,Myr, however, due to the rapid rotation and possible oblate spheroid nature of the primary, the photometric prescriptions for determining the effective temperature and reddening of the primary used by \citet{pecaut12} are likely to fail for this object. The spectral type is more consistent with a temperature of $\sim$30000\,K. 
Additionally, the presence of other evolved B-type stars is evidence for an older population \citep{pecaut12}. Furthermore, the recent age estimate of 13\,Myr for the F-type members of Upper Scorpius by \citet{pecaut12}  further supports an older population in the subgroup. If the HR diagram position on EW(Li) that we observe among our members is real than this also supports multiple age population in Upper Scorpius.

\begin{figure}
\subfloat[\label{wjkw2}]{\includegraphics[width=0.47\textwidth]{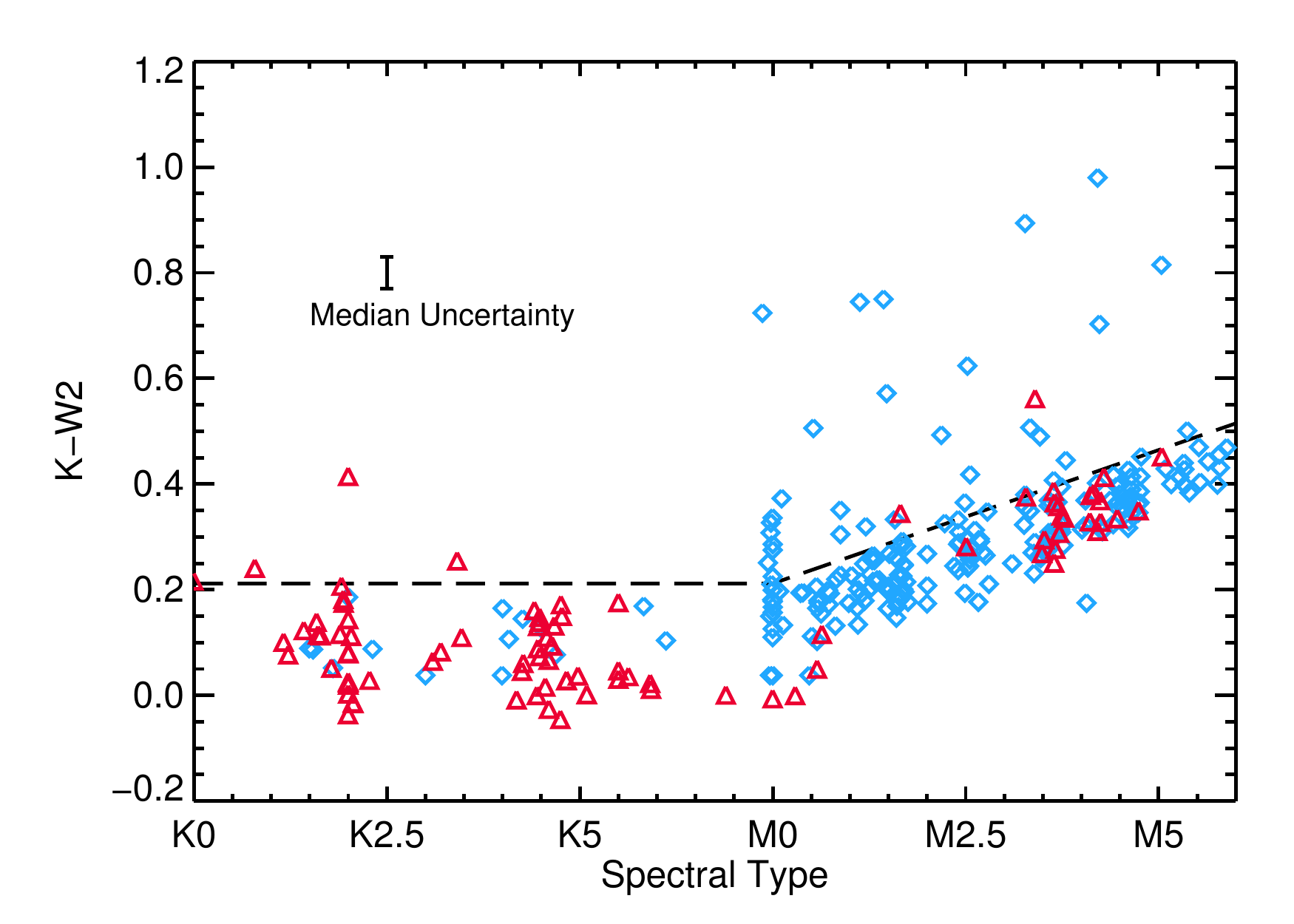}}\\
\subfloat[\label{wjkw3}]{\includegraphics[width=0.47\textwidth]{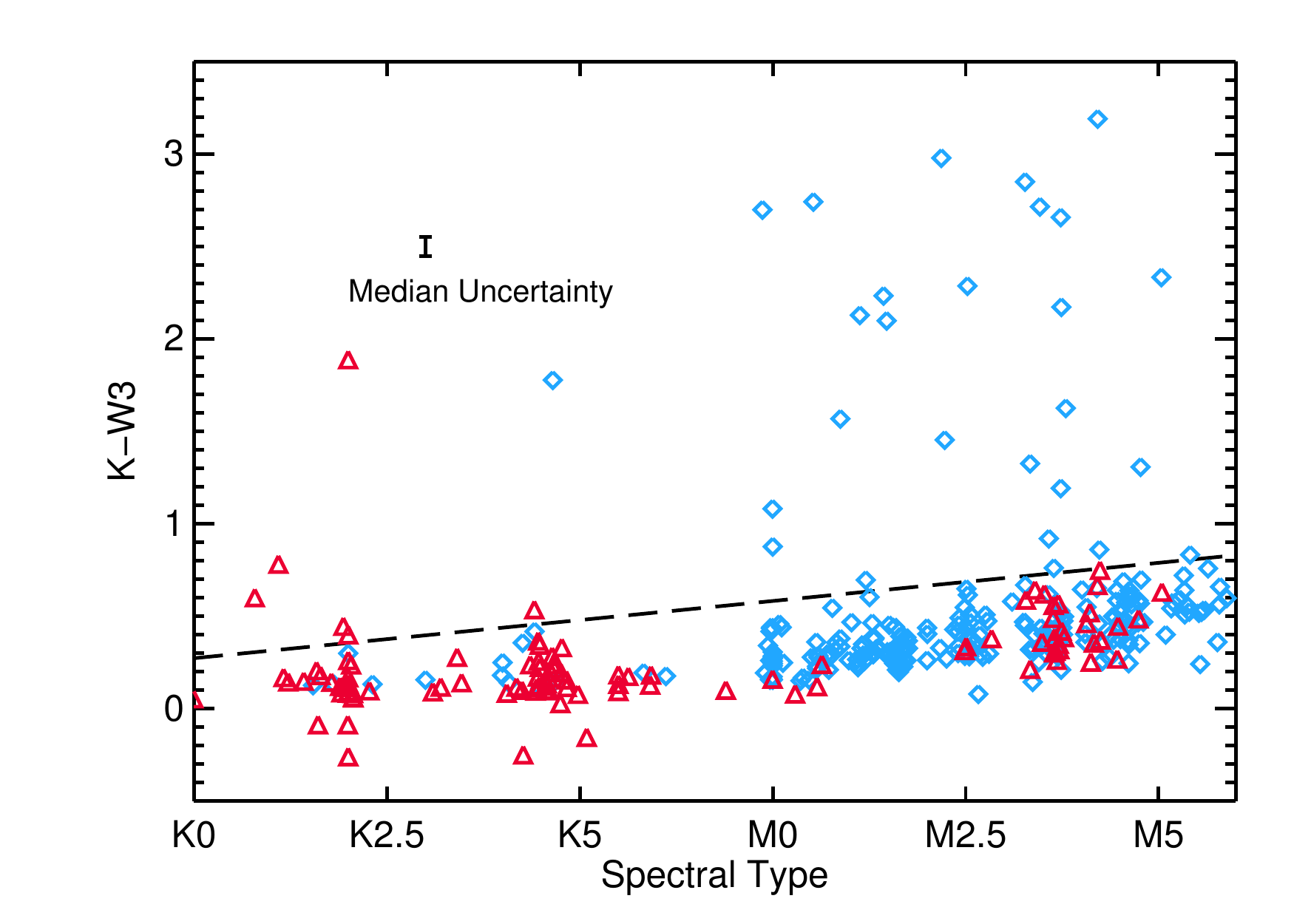}}\\
\subfloat[\label{wjkw4}]{\includegraphics[width=0.47\textwidth]{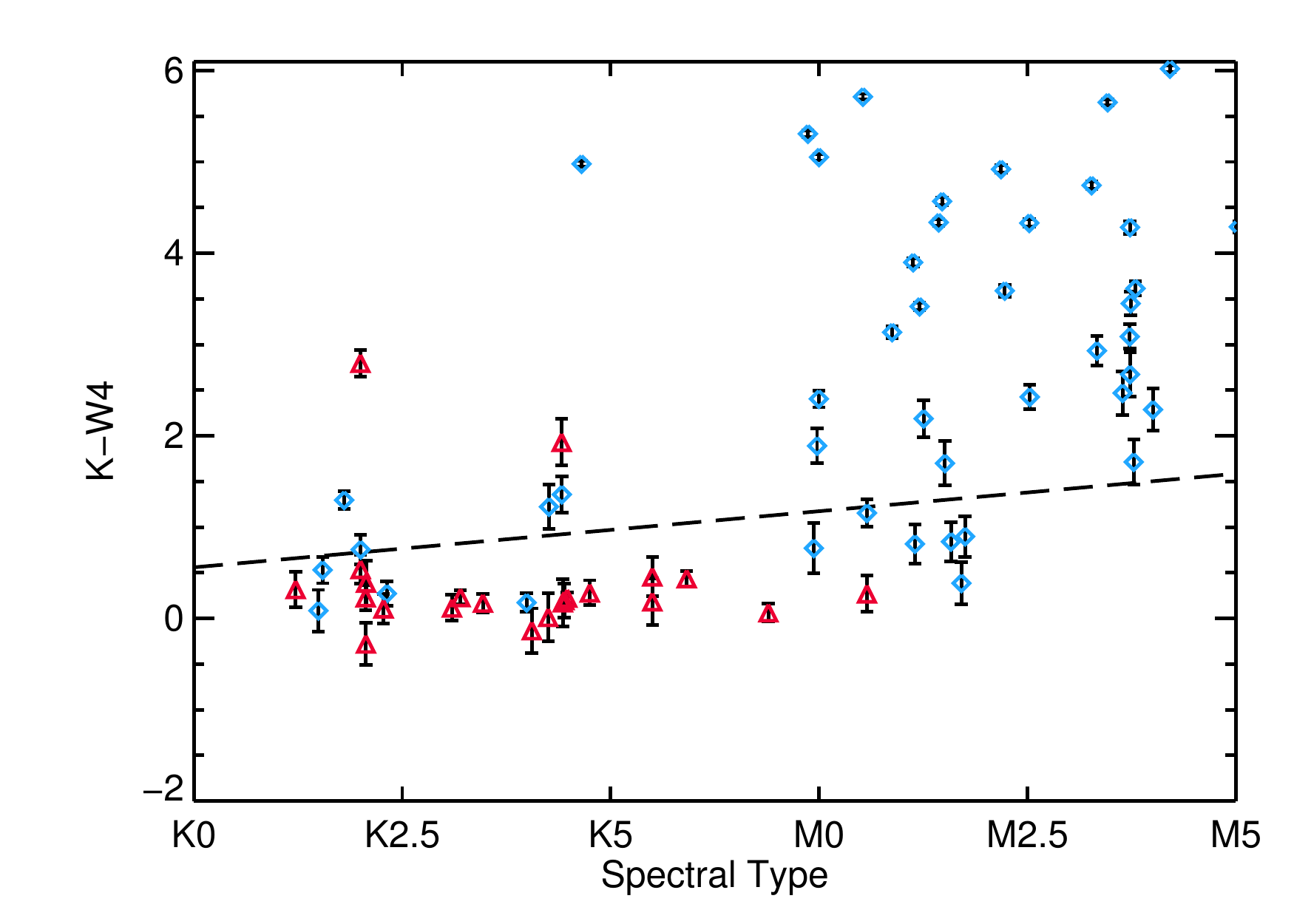}}
\caption{Near IR and WISE band color-color diagrams for both the newly identified members (blue diamonds) and non-members (red triangles) from our spectroscopic survey for $(J-K, K-W2)$, $(J-K, K-W3)$, and $(J-K, K-W4)$. We have omitted objects in each WISE band that were flagged as having poor or contaminated photometry in the catalog. The dashed lines indicate the position of the upper boundary of the photospheric sequence. If a star is above this threshold we designate it as displaying an excess in the particular band.}
\label{excess_figs}
\end{figure}

\section{Disk Candidates}

We have also obtained the Wide-Field Infrared Survey Explorer (WISE) infrared photometry \citep{wise10}, from the ALLWISE version of the catalog, for the observed candidate members in order to determine the prevalence of circumstellar disks among our new members. The identification of new populations of stars bearing disks is valuable because it provides extension to the current samples used in the study of disk property measurements and disk evolution. The AllWISE catalog provides photometry in four bands W1, W2, W3 and W4, with effective wavelengths of 3, 4.5, 12 and 22\,$\mu$m respectively. The W2 and W3 photometry is effective for tracing the presence of an inner disk, while excess in the W4 band photometry can indicate the presence of a colder, outer disk or transitional disk.

We queried the ALLWISE catalog for the positions of the 397 stars we observed from our sample, including 237 new members, with a search radius of 5". The search returned 395 matches with varying levels of photometric quality. We then placed each star on three spectral type-color diagrams incorporating 2MASS \citep{2mass} K-band photometry, these were K-W2, K-W3, and K-W4. Past studies have used both K, and W1 as the base photometry for building color-color diagrams \citep{carpenter06,carpenter09, rizzuto12, luhman10, luhman2012_disk}. Typically, the presence of a disk within $\sim3$\,AU of a host star increases the brightness in the IR wavelengths, with $\sim$5\,$\mu$m being the approximate wavelength where the disk dominates in brightness. Both the W1 and K magnitudes are long enough such that reddening is not a significant issue, but also shorter than the expected point of disk domination. We found that examining the WISE bands relative to the K magnitude produced a better separation of disk bearing stars from photospheric emission, and so we report the analysis in terms of this methodology. 

Figure \ref{excess_figs} displays the three spectral type-color diagrams. We excluded any WISE photometry in a given band that was flagged as having a signal to noise ratio of $<4$, as a non-detection, or flagged as being contaminated by any type of image artifact in the catalog. This resulted in the exclusion of 56, 10 and 312 objects in the W2, W3, and W4 bands respectively. The primary source of the exclusions for the W4 band was non-detection or low signal to noise at 22\,$\mu$m, and most of the exclusions in the W2 and W3 bands were due to contamination by image artifacts. To reduce contamination by extended sources we also excluded any object flagged as being nearby a known extended source or with significantly poor photometry fits, there were eight such objects. The WISE band images for these stars were then inspected visually to gauge the extent of contamination. We found the three of the objects were not significantly effected by the nearby extended source, and so included them in the analysis. After excluding these objects we were left with 333, 379 and 77 objects with photometry of sufficient quality in the W2, W3 and W4 bands respectively. 

Due to the age of Upper Scorpius of $\sim$10\,Myr, the majority of members no longer possess a disk, providing sufficient numbers of stars to clearly identify photospheric emission. Hence the photosphere color can be determined from the clustered sequence in the spectral type-color diagrams. We fit a straight lines in the K-W3 and K-W4 WISE band colors, and a disjointed line in K-W2, and then place a boundary where the photospheric sequence ends. For K-W3 the boundary line is given by the points (K0,0.27) and (M5,0.8) and for K-W4 the points (K0,0.56) and (M5,1.6).The sloped part of the boundary line for K-W2 is defined by the points (M0,0.21) and (M5,0.46), and the flat section by K-W2 $=0.21$, for spectral types earlier than M0. These boundaries are shown as black lines in Figures \ref{excess_figs}. Stars with color redder than these boundaries we deem to display an excess in the particular WISE band. Upon inspection, we find that our placement of the end of the photospheric sequence is closely consistent with that of \citep{luhman2012_disk}. In the K-W4 color, we find that for stars of spectral type later than $\sim$M2, the photospheric emission in W4 is undetectable by WISE.

For those stars which displayed excesses in any combination of WISE bands, we visually inspected the images to exclude the possibility that the excesses could have been caused by the presence of close companions or nebulosity. We also found that in a few cases background structure in the W4 image could cause the appearance of an excess, although this effect was largely mitigated by our signal-to-noise cutoff. We rejected 23 of the excess detections after inspection, 12 of which were caused by background structure or nearby nebulosity, and 11 of which were due to blending with nearby stars. We further excluded any object which shows an excess in only the K-W2 color as likely being produced by unresolved multiplicity. After these rejections, 27 stars remained with reliable excess detections. Additionally, a single object, 2MASS J16194711-2203112 , displayed an excess in K-W3, but had a W4 detection with signal-to-noise of 3.5. Upon inspection of the corresponding W4 band image, we included it as exhibiting an excess in K-W4.

\begin{table*}
\begin{tabular}{ccccccccc}
\hline
& R.A. & Decl. &  &  & & & & \\
2MASS & (J2000.0) & (J2000.0) & M & E & D & W2 & W3 & W4 \\
\hline
15581270-2328364 & 15 58 12.70 &  -23 28 36.4 & Y & XXE & D/ET & 7.97&7.88&6.72\\
15573430-2321123 & 15 57 34.31 &  -23 21 12.3 & Y & EEE & E & 8.67&8.30&5.58\\
16052157-1821412 & 16 05 21.57 &  -18 21 41.2 & Y & XEE & T & 7.24&6.36&3.16\\
16064385-1908056 & 16 06 43.86 &  -19 08 05.5 & Y & EEE & E & 8.86&8.11&6.79\\
16064794-1841437 & 16 06 47.94 &  -18 41 43.8 & Y & XEE & T & 8.75&8.10&3.93\\
16120668-3010270 & 16 12 06.68 &  -30 10 27.1 & Y & EEE & F & 8.81&6.58&3.60\\
16134781-2747340 & 16 13 47.82 &  -27 47 34.0 & Y? & XEE & E & 7.93&7.81&7.04\\
15594426-2029232 & 15 59 44.27 &  -20 29 23.4 & Y & EEE & F & 9.59&8.07&6.12\\
16120505-2043404 & 16 12 05.05 &  -20 43 40.5 & Y & EEE & E & 8.71&7.50&5.93\\
16112601-2631558 & 16 11 26.03 &  -26 31 55.9 & Y & XEE & E & 9.24&8.12&5.98\\
16093164-2229224 & 16 09 31.66 &  -22 29 22.4 & Y & EEE & F & 8.65&6.17&4.23\\
16252883-2607538 & 16 25 28.81 &  -26 07 53.8 & Y & XXE & D/ET & 9.61&9.24&7.43\\
16333496-1832540 & 16 33 34.97 &  -18 32 53.9 & Y & XEE & E & 9.81&9.21&7.72\\
16023587-2320170 & 16 02 35.89 &  -23 20 17.1 & Y & XEE & E & 9.79&8.31&7.41\\
16011398-2516281 & 16 01 13.99 &  -25 16 28.2 & Y & EEE & E & 9.97&8.79&6.81\\
16132190-2136136 & 16 13 21.91 &  -21 36 13.7 & Y & EEE & F & 9.40&7.88&5.41\\
16194711-2203112 & 16 19 47.12 &  -22 03 11.2 & Y & XEX & E & 10.13&9.24& \\
16041893-2430392 & 16 04 18.93 &  -24 30 39.3 & Y & EEE & F & 8.23&6.57&4.52\\
16100501-2132318 & 16 10 05.02 &  -21 32 31.9 & Y & EEE & F & 8.23&6.25&3.64\\
16200616-2212385 & 16 20 06.16 &  -22 12 38.5 & Y & XEE & E & 10.12&8.48&7.20\\
15564244-2039339 & 15 56 42.45 &  -20 39 34.2 & Y & EEE & F & 9.79&7.57&4.63\\
16271273-2504017 & 16 27 12.74 &  -25 04 01.8 & Y & EEE & F & 8.64&7.25&5.48\\
15583620-1946135 & 15 58 36.20 &  -19 46 13.5 & Y & EEE & F & 9.74&7.53&4.70\\
16012902-2509069 & 16 01 29.03 &  -25 09 06.9 & Y & EEE & F & 9.19&7.24&5.34\\
16071403-1702425 & 16 07 14.02 &  -17 02 42.7 & Y & XEE & F & 10.25&8.10&6.47\\
16145244-2513523 & 16 14 52.41 &  -25 13 52.4 & Y & EEE & E & 9.95&9.13&7.52\\
16153220-2010236 & 16 15 32.20 &  -20 10 23.7 & Y & EEE & F & 8.16&6.68&4.57\\

\hline
\end{tabular}
\caption{List of stars in our sample with observed IR excesses in the WISE bands. (M) indicates the membership status of the star, (E) list the detection of IR excesses in the three WISE bands, (D) labels the candidate disk classification where F, E, T and D/ET mean full, evolved, transition, and debris or evolved transition respectively. The final three columns list the availability of photometry in the three longest WISE bands.  The question mark (?) indicates the star HD-145778, which is likely to be an F-type members of Upper Scorpius.}
\label{excess_table}
\end{table*}

\begin{figure}
\subfloat[\label{wk4wk2}]{\includegraphics[width=0.47\textwidth]{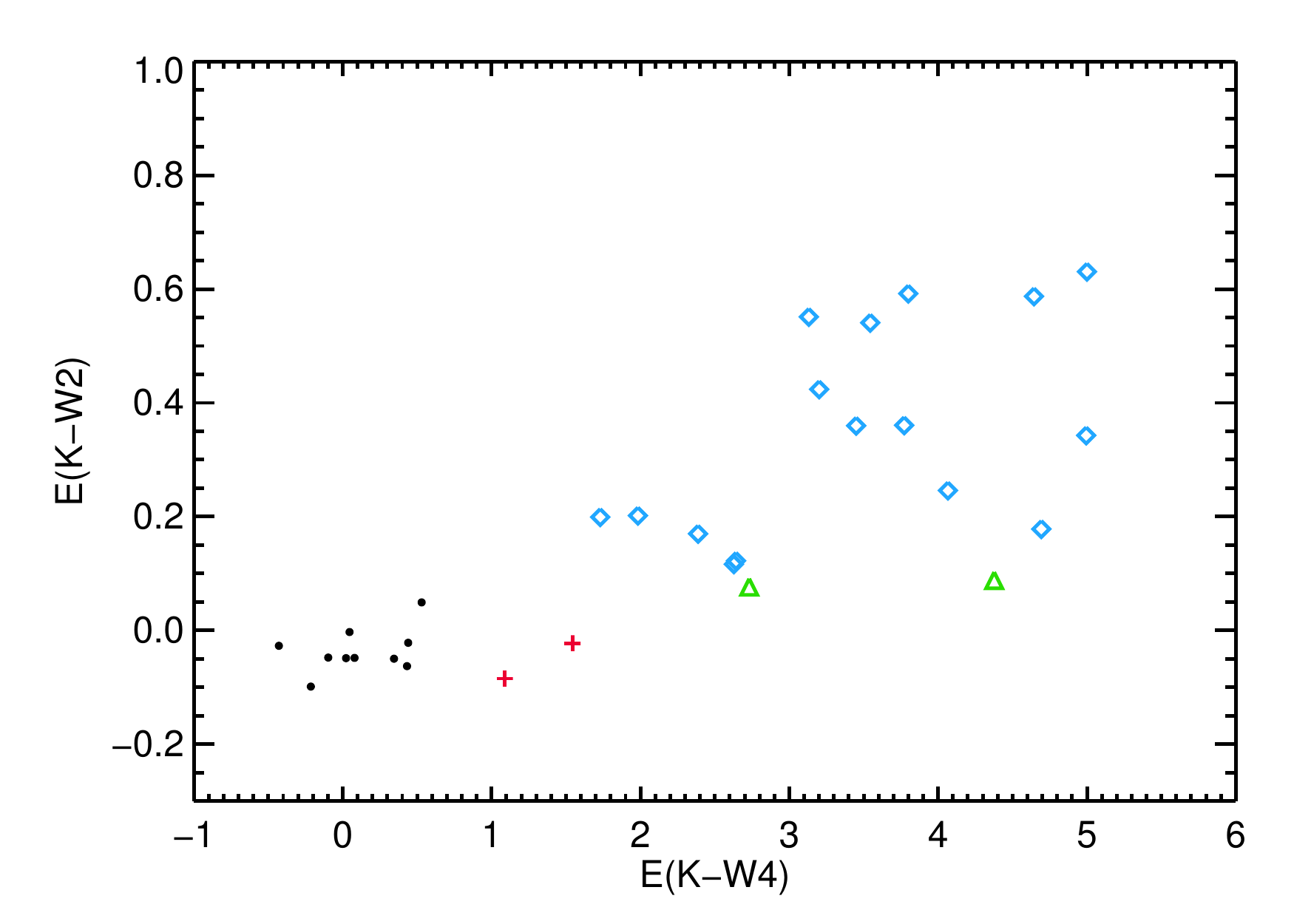}}\\
\subfloat[\label{wk4wk3}]{\includegraphics[width=0.47\textwidth]{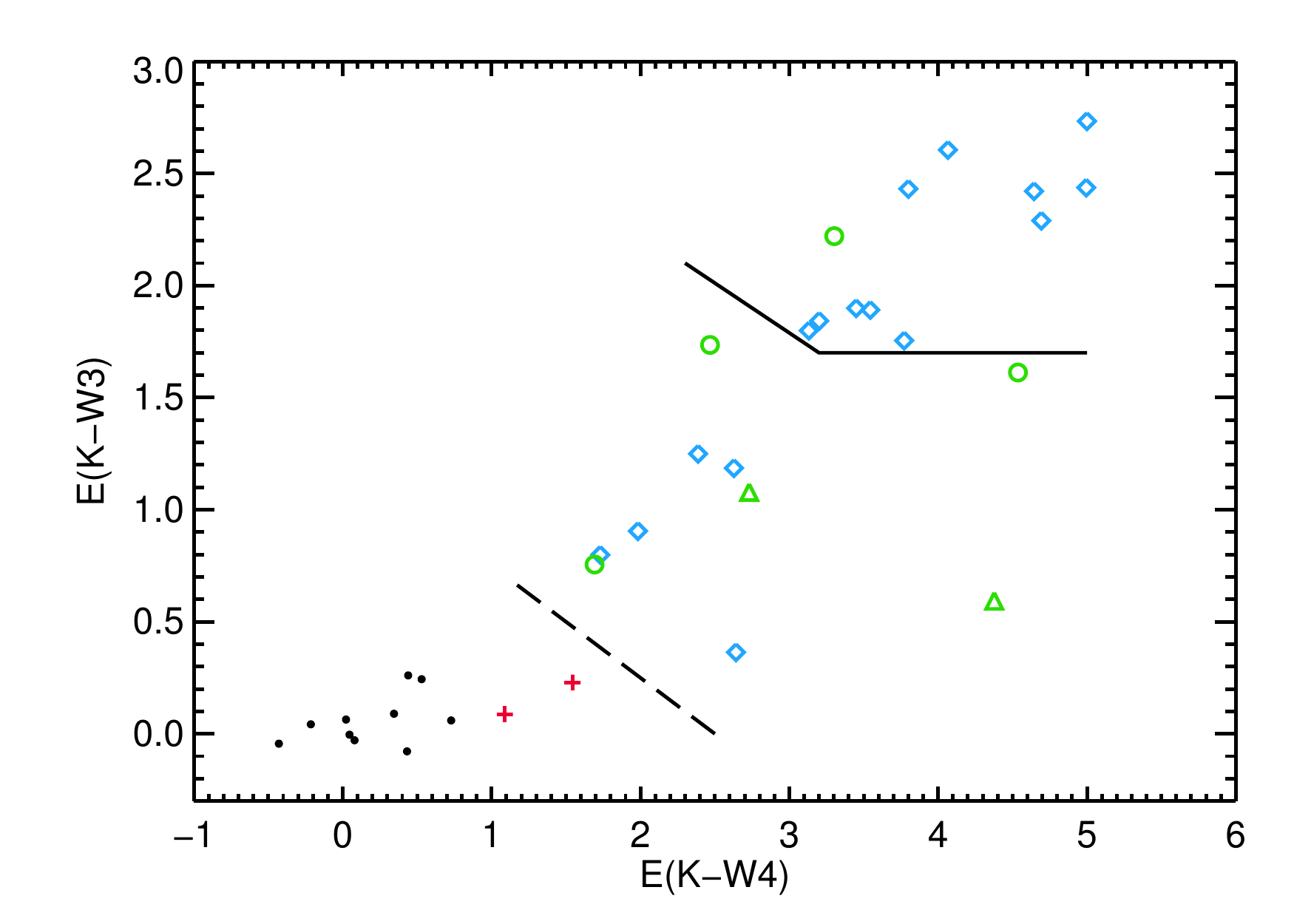}}
\caption{Excesses in K-W2 vs. K-W4 and K-W3 vs. K-W4 for the members observed in our survey. The plots includes members with out excess (black circles), with excesses in W2, W3 and W4 (blue diamonds), with excesses in the two longest WISE bands with and without reliable W2 photometry (green triangles and circles respectively),  and members with an excess in only the W4 band (red plusses). We also include three members with just W3 excesses, which were detected in W4, but without excesses which clearly sit above the diskless sequence (purple squares). The black lines indicate adopted boundaries for different disk classifications.}
\label{wk4wk}
\end{figure}

We can classify the disk types by the amount of excess displayed in different colors compared to the photosphere. We adopt disk type criteria in the E(K-W4), E(K-W3) space consistent with those described in \citet{luhman10} and \citet{luhman2012_disk}, which identify four different categories of disk: Full or primordial disks, transition disks, evolved disks, and debris or evolved transition disks. Primordial or full disks exhibit strong emission across the entire IR spectral range. Transition disks are structurally different in that they have a significant cleared inner hole, which is visible as a weaker emission at the shorter IR wavelengths, but still relatively bright in the longer IR wavelengths. Evolved disks do not show a gap in IR emission, but have started to become thinned and appear fainter at all IR wavelengths than unevolved full disks, with a steady decline in IR excess with age \citep{carpenter09}. Debris disk and evolved transition disk have similar IR SED's, showing only weak excesses at the longer IR wavelengths. Figure \ref{wk4wk3} show both E(K-W4) and   E(K-W3) for the stars identified as having displaying an excess. The lines in Figure \ref{wk4wk3} bound the different regions populated by the various disk types. We classify all objects with excesses in W3 and W4 beneath the dashed line to be debris or evolved transition disks candidates, and the objects above the solid line to be full disks. Stars with excess between these two lines we classify as evolved disk candidates. Finally, we identify the two objects with a large W4 excess, but W3 excesses too small to be classified as full disks, as transition disk candidates. Table \ref{excess_table} lists the excess status for the stars with detected excesses. 

In total, we identify 26 of the Upper Scorpius members as displaying a disk-indicating excess with spectral types later than K0, and  one star without significant Lithium absorption that also displays an excess. This latter object is an F4.5 spectral type object,, HD-145778, with $EW(Li)=0.09\pm0.2$. The presence of some Lithium absorption, combined with the disk presence mean that this object can be considered to be a member of Upper Scorpius. We have included it as a member at the end of Table \ref{obs_res_tab}. HD-145778 is not in this HIPPARCOS catalog \citep{lindegren97}, potentially explaining why it was not included in past memberships.

Due to the WISE detection limit in the W4 band, we are almost certainly not able to identify the vast majority of the evolved transitional and debris disks, which show only a small color excess in K-W4. Indeed, we only detect two such disks in our sample, one of which, USco 41, was previously identified with Spitzer photometry \citep{carpenter09}, when significantly more are expected from previous statistics \citep{carpenter09, luhman2012_disk}.  Furthermore, it is likely that a number of evolved disks around stars of spectral type later than $\sim$M3 are not detected here. For this reason it is difficult to meaningfully estimate the disk or excess fraction for our entire sample. In the M0 to M2 spectral type range, where we expect the majority of the full, evolved and transitional disks to be detectable by WISE, we have 11 disks, 6 of which are full, 4 evolved and 1 transitional. Excluding all those members flagged for extended emission, confusion with image artifacts, or unreliable excesses, we find and excess fraction of 11.2$\pm$3.4\%. \citet{carpenter09} found a primordial disk fraction for M-type Upper Scorpius members of $\sim$17\%, and \citet{luhman2012_disk} find excess fractions of 12\% and 21\% for K-type and M0 to M4-type members respectively. Given the strong increase in excess fraction towards the late M-type members and the potential for some missed evolved disks due to the WISE detection limits, we find that our excess fraction estimate is consistent with these past results. 

\section{Conclusions}

We have conducted a spectroscopic survey of 397 candidate Upper Scorpius association K and M-type members chosen through statistical methods, and revealed 237 new PMS members among the sample based on the presence of Li absorption. We also identify 25 members in our sample with WISE near-infrared excesses indicative of the presence of a circumstellar disk, and classify these disk on the basis of their color excess in different WISE bands. We find that the members show a significant spread in EW(Li), and upon placing the members on a HR diagram, we find that there is a potential age spread, with a small correlation between EW(Li) and HR-diagram position. This could indicate the presence of a distribution of ages, or multiple populations of different age in Upper Scorpius.

\bibliography{master_reference}
\bibliographystyle{mn2e}

\appendix
\onecolumn
\section{Bayesian Selection of Candidate PMS Upper Scorpius Stars}
\label{bayesapp}

The Bayesian selection method we have employed to identify candidate Upper Scorpius members in the low-mass (K and M-type) regime is largely based upon the high-mass star selection of \citet{myfirstpaper}, with some significant changes to accommodate the particular data available for the low-mass stars. 
As in \citet{myfirstpaper} we consider two models: (1) The Upper Scorpius group model ($M_g$) and (2) the field model describing ordinary field stars ($M_f$). These models both provide a variety of information, including position, distance and velocity distributions. For use with the low-mass stars, we also include a model isochrone;

\begin{equation}
M_{g,f}(l,C_x) = \{l,b,r,U,V,W,M_x\},
\label{models}
\end{equation}

where $l$ and $b$ are Galactic longitude and latitude, $r$ is distance, $U, V$ and $W$ are the three components of a star's Galactic velocity, $C_x$ represents various colours, and $M_x$  absolute magnitudes in corresponding filters. The model values of distance and velocity are dependent on the Galactic longitude of a candidate star, while the model absolute magnitude is dependent on the star's colour. For the application to PMS stars in the Sco-Cen subgroups, we have used Siess isochrones \citep{siess00} of 6\,Myr for US and 16\,Myr for UCL and LCC  for the group models ($M_g$) and an older main-sequence isochrone for the field model ($M_f$).

For the kinematic models, we employ the same linear models in Galactic longitude for the group model, and the Galactic thin disk model of \citep{robin03}, as were used in \citet{myfirstpaper}.

We then calculate the model likelihood ratio for these two model for each stars according to equations 10 and 11 of \citep{myfirstpaper};

\begin{equation}
R = \frac{P(M_g|D)}{P(M_f|D)}  = \frac{P(M_g)}{P(M_f)} \frac{\int P(D|\boldsymbol{\phi}_g,M_g)P(\boldsymbol{\phi}_g|M_g)\mathrm{d}\boldsymbol{\phi}_g}{\int P(D|\boldsymbol{\phi}_f,M_f)P(\boldsymbol{\phi}_f|M_f)\mathrm{d}\boldsymbol{\phi}_f},
\label{bfact_final}
\end{equation}

where again, $M_{g,f}$ represents the association and field models, and $\phi_{g,f}$ represents the set of parameters, or a parameter vector derived from the models, which can be directly compared to the data $D$.  The data are compiled from positions and proper motions taken from the UCAC4 catalog \citep{ucac4} and any available radial velocities in the RAVE catalog \citep{ravedr1},and a photometric distance calculated from the APASS B and V band photometry \citep{apass} and 2MASS J,H and K photometry \citep{2mass}, using the model isochrones. Included in our distance estimate was a multiplicity photometric bias correction. This estimate was based on the expected multiplicity statistics of G, K and M-type stars, taken from \citet{kraus11}, which indicates $\sim$50\% of solar type stars have a companion. Combined with a standard initial mass function for the companion, this produces an average multiplicity bias of 0.2\,magnitudes. Our photometric distances are thus calculated by adjusting the measured photometry by 0.2 magnitudes and then calculating a distance based on an interpolated isochrone magnitude. We define the uncertainty on the photometric distance conservatively to be 20\%, or approximately $\pm$0.4 magnitudes. This calculation is done for every star in our sample for both the association and field models.

\begin{figure}
\centering
\includegraphics[width=0.6\textwidth]{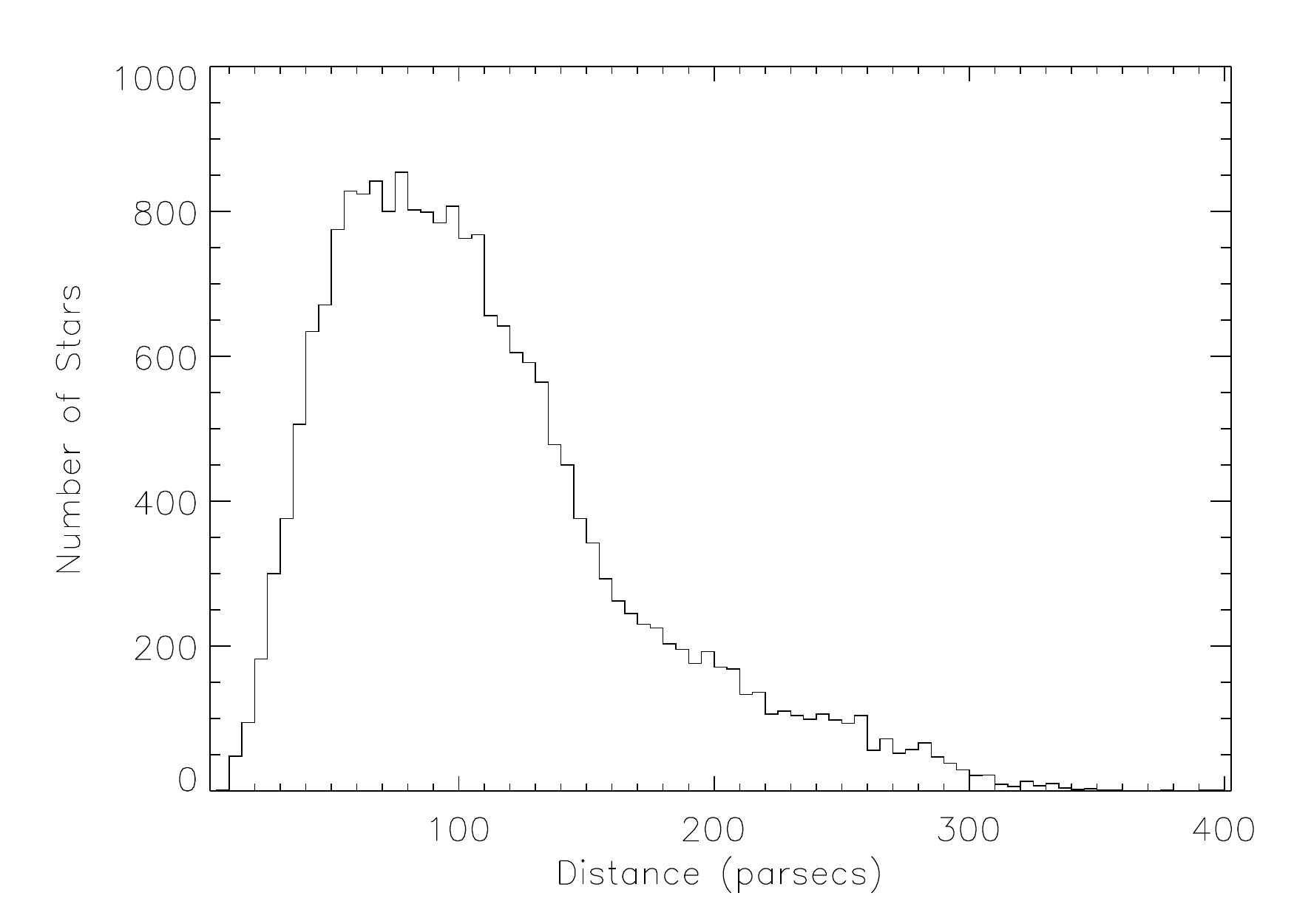}
\caption{Field photometric distance distribution, showing features of the input catalog magnitude and colour cuts, and the isochrone used in estimation of the distances.}
\label{field_dist_fig}
\end{figure}

With the inclusion of the photometric distances, we then calculate the model likelihood ratio integrals in the same way as described in \citet{myfirstpaper} with a difference only present in how the distance and proper-motion integrals are treated. Equations \ref{prob_pars_mod}, \ref{groupdist} and \ref{prob_dat_pars} show the details of the terms in the integrals of equation \ref{bfact_final}. For the high-mass stars, for which relatively well-defined parallax measurements were available, we treated the distance and parallel proper motion (proper motion in the direction of the association movement) integrals separately. For the larger distance and proper motion uncertainties of the low-mass stars, the two dimensional probability distribution in proper motion-parallax space is not symmetrical. This means that separating the proper motion and parallax integrals would inadequately describe the true value of the two dimensional integral. We address this issue by first sampling from the model distance distribution and then comparing the sampled values to the photometric distance measures and proper motion. We take $10^5$ and $10^6$ random samples of $(U,V,W,r)_{g,f}$ for the group and field models respectively to calculate the integrals. More samples are required for adequate sampling of the field model due to the larger spread of possible values. The key difference here when compared to the high-mass selection of \citet{myfirstpaper} is that we also sample random distances from the distributions. The following equation describes the velocity distributions for both the field and the group models;

\begin{equation}
P(\boldsymbol{\phi}_{g,f}|M_{g,f}) \propto \exp{\left(-\frac{(U-U_{g,f}(l))^2}{2\sigma^2_{U_{g,f}}} \right) +  \left(-\frac{(V-V_{g,f}(l))^2}{2\sigma^2_{V_{g,f}}} \right) + \left(-\frac{(W-W_{g,f}(l))^2}{2\sigma^2_{W_{g,f}}} \right)}.
\label{prob_pars_mod}
\end{equation}

where $U$, $V$, $W$ and $r$ represent the three components of the Galactic velocity and distance, with subscripts $g$ and $f$ indicating that the values are the corresponding expected values for the two models. The distance distribution for the group is also taken to be a normal distribution, as in \citet{myfirstpaper};
\begin{equation}
\exp{\left(- \frac{(r-r_{g,f}(l))^2}{2\sigma^2_{r_{g,f}}} \right)}
\label{groupdist}
\end{equation}
where $r$ is the sampled distance and $r_{g,f}$ is the expected distance. Note that the model distribution means are a function of the Galactic longitude of the star being examined. For the field model, the distance distribution is not as directly describable as that of the group model. Instead, we construct the field distribution by taking all the calculated photometric distances for the field model and then take the samples from this using rejection sampling. Figure \ref{field_dist_fig} shows the field distribution used in the sampling for the field model. 

The final term in equation \ref{bfact_final} is then given by;
\begin{equation}
P(D|\boldsymbol{\phi}_{g,f}) \propto \exp{\left(-\frac{\mu_\perp^2}{2\sigma^2_{\mu_{\perp}}} 
-\frac{(\nu_r - \nu_{r_{g,f}})^2}{2\sigma^2_{\nu_r}}
-\frac{(\mu_\parallel - \frac{\mu_{\parallel_{g,f}}}{r} )^2}{2\sigma^2_{\mu_\parallel}} 
-\frac{(\pi_{g,f}- \frac{1}{r})^2}{2\sigma^2_{\pi_{g,f}}}\right)},
\label{prob_dat_pars}
\end{equation}

where $\mu_\parallel$ and $\mu_\perp$ are the star proper motions in the direction parallel to and perpendicular to the expected direction of motion for a given set of $(U,V,W)$. For more information regarding this coordinate system, including procedures for transforming to and from equatorial coordinates, see \citet{zeeuw99} and \citet{myfirstpaper}. $\pi_{g,f}$ is the calculated photometric parallax for the group and field models respectively,  and $\nu_r$ is the radial velocity, which in the vast majority of cases is unconstrained and does not influence the integrals. With the model likelihood ratio we then determine the probability of membership, given by $R/(R+1)$.



\section*{Supplementary material}

| | | | | | | | | | |
|---|---|---|---|---|---|---|---|---|---|
| RIK-55  | J16010896-2645104 | 16 01 08.97 | -26 45 10.6 | 0.47 | 0.05 | -5.38    | 0.19  | M3.5 | 0.0 |
| RIK-56  | J16011398-2516281 | 16 01 13.99 | -25 16 28.2 | 0.48 | 0.04 | -5.88    | 0.15  | M4.0 | 0.0 |
| RIK-57  | J16012233-1937222 | 16 01 22.34 | -19 37 22.3 | 0.45 | 0.02 | -2.59    | 0.03  | M1.5 | 0.7 |
| RIK-58  | J16012902-2509069 | 16 01 29.03 | -25 09 06.9 | 0.34 | 0.03 | -129.46  | 13.94 | M3.5 | 0.3 |
| RIK-59  | J16015987-1843457 | 16 01 59.88 | -18 43 45.7 | 0.36 | 0.02 | -4.25    | 0.10  | M2.5 | 0.9 |
| RIK-60  | J16020039-2221237 | 16 02 00.39 | -22 21 23.9 | 0.66 | 0.04 | -4.23    | 0.05  | M1.0 | 0.7 |
| RIK-61  | J16020845-2254588 | 16 02 08.45 | -22 54 59.1 | 0.67 | 0.03 | -3.50    | 0.05  | M1.0 | 0.6 |
| RIK-62  | J16021489-2438325 | 16 02 14.89 | -24 38 32.6 | 0.66 | 0.09 | -15.33   | 0.46  | M5.5 | 0.7 |
| RIK-63  | J16022357-2259332 | 16 02 23.57 | -22 59 33.3 | 0.49 | 0.09 | -13.91   | 1.23  | M6.0 | 1.0 |
| RIK-64  | J16022461-2200248 | 16 02 24.61 | -22 00 24.8 | 0.72 | 0.04 | -3.92    | 0.07  | M1.5 | 0.0 |
| RIK-65  | J16023587-2320170 | 16 02 35.89 | -23 20 17.1 | 0.53 | 0.03 | -8.46    | 1.34  | M3.5 | 0.6 |
| RIK-66  | J16024448-2543323 | 16 02 44.48 | -25 43 32.3 | 0.45 | 0.04 | -8.79    | 1.50  | M4.0 | 0.0 |
| RIK-67  | J16030177-2626218 | 16 03 01.77 | -26 26 21.9 | 0.46 | 0.03 | -2.95    | 0.15  | M0.0 | 0.0 |
| RIK-68  | J16031491-2234454 | 16 03 14.91 | -22 34 45.5 | 0.48 | 0.05 | -15.50   | 9.06  | M4.5 | 0.3 |
| RIK-69  | J16032599-2627320 | 16 03 25.98 | -26 27 32.2 | 0.83 | 0.07 | -9.27    | 0.11  | M4.5 | 0.4 |
| RIK-70  | J16033777-1845083 | 16 03 37.77 | -18 45 08.3 | 0.56 | 0.04 | -4.65    | 0.08  | M1.0 | 0.5 |
| RIK-71  | J16033829-1854076 | 16 03 38.30 | -18 54 07.7 | 0.55 | 0.02 | -4.37    | 0.04  | M1.0 | 0.8 |
| RIK-72  | J16033922-1851294 | 16 03 39.22 | -18 51 29.3 | 0.59 | 0.03 | -2.28    | 0.14  | M2.5 | 0.6 |
| RIK-73  | J16034695-2245246 | 16 03 46.95 | -22 45 24.7 | 0.42 | 0.04 | -3.86    | 0.31  | M1.5 | 0.0 |
| RIK-74  | J16035175-2140154 | 16 03 51.75 | -21 40 15.5 | 0.55 | 0.04 | -14.20   | 0.64  | M6.0 | 0.7 |
| RIK-75  | J16035404-2509393 | 16 03 54.05 | -25 09 39.4 | 0.59 | 0.06 | -7.55    | 0.32  | M5.5 | 0.2 |
| RIK-76  | J16040671-2637070 | 16 04 06.71 | -26 37 07.1 | 0.52 | 0.03 | -4.27    | 0.04  | M1.5 | 0.6 |
| RIK-77  | J16041893-2430392 | 16 04 18.93 | -24 30 39.3 | 0.57 | 0.02 | -24.08   | 0.37  | M2.5 | 0.3 |
| RIK-78  | J16052157-1821412 | 16 05 21.57 | -18 21 41.2 | 0.47 | 0.02 | -37.96   | 0.15  | K4.5 | 1.0 |
| RIK-79  | J16053815-2039469 | 16 05 38.16 | -20 39 47.0 | 0.59 | 0.02 | -1.09    | 0.03  | M0.0 | 0.3 |
| RIK-80  | J16064385-1908056 | 16 06 43.86 | -19 08 05.5 | 0.59 | 0.02 | -8.57    | 0.07  | M0.0 | 0.6 |
| RIK-81  | J16064794-1841437 | 16 06 47.94 | -18 41 43.8 | 0.49 | 0.02 | -0.92    | 0.04  | M0.0 | 0.3 |
| RIK-82  | J16071403-1702425 | 16 07 14.02 | -17 02 42.7 | 0.71 | 0.04 | -3.70    | 0.09  | M3.5 | 0.7 |
| RIK-83  | J16073037-2546269 | 16 07 30.38 | -25 46 27.0 | 0.44 | 0.01 | -2.82    | 0.03  | M1.5 | 0.5 |
| RIK-84  | J16074006-2148426 | 16 07 40.06 | -21 48 42.7 | 0.51 | 0.02 | -2.06    | 0.05  | M0.5 | 0.3 |
| RIK-85  | J16075136-1718231 | 16 07 51.37 | -17 18 23.2 | 0.54 | 0.03 | -4.09    | 0.04  | M2.5 | 0.6 |
| RIK-86  | J16080141-2027416 | 16 08 01.42 | -20 27 41.7 | 0.62 | 0.08 | -1.93    | 0.18  | M0.0 | 0.9 |
| RIK-87  | J16080411-2640449 | 16 08 04.11 | -26 40 44.9 | 0.47 | 0.06 | -7.81    | 0.54  | M0.5 | 0.0 |
| RIK-88  | J16082078-2131234 | 16 08 20.77 | -21 31 23.4 | 0.70 | 0.04 | -5.62    | 0.19  | M4.5 | 0.7 |
| RIK-89  | J16085628-2148489 | 16 08 56.29 | -21 48 48.9 | 0.56 | 0.02 | -3.71    | 0.03  | M1.5 | 0.5 |
| RIK-90  | J16085695-2835573 | 16 08 56.96 | -28 35 57.4 | 0.21 | 0.06 | -10.18   | 0.18  | M4.5 | 0.9 |
| RIK-91  | J16090051-2745194 | 16 09 00.52 | -27 45 19.4 | 0.80 | 0.18 | -11.25   | 0.62  | M5.5 | 1.1 |
| RIK-92  | J16092063-2222057 | 16 09 20.63 | -22 22 05.7 | 0.56 | 0.08 | -16.49   | 0.39  | M5.5 | 0.8 |
| RIK-93  | J16092969-2200579 | 16 09 29.70 | -22 00 58.0 | 0.59 | 0.02 | -3.34    | 0.03  | M1.5 | 0.7 |
| RIK-94  | J16093030-2104589 | 16 09 30.31 | -21 04 58.9 | 0.49 | 0.02 | -0.76    | 0.03  | M0.0 | 0.3 |
| RIK-95  | J16093108-2041460 | 16 09 31.10 | -20 41 46.0 | 0.53 | 0.03 | -9.06    | 0.06  | M4.5 | 1.0 |
| RIK-96  | J16093164-2229224 | 16 09 31.66 | -22 29 22.4 | 0.38 | 0.07 | -4.90    | 0.10  | M2.0 | 0.9 |
| RIK-97  | J16093575-2138057 | 16 09 35.76 | -21 38 05.8 | 0.55 | 0.01 | -3.66    | 0.03  | M2.5 | 0.7 |
| RIK-98  | J16093969-2200466 | 16 09 39.70 | -22 00 46.6 | 0.51 | 0.03 | -3.00    | 0.08  | M0.5 | 0.7 |
| RIK-99  | J16094955-2444468 | 16 09 49.55 | -24 44 46.8 | 0.55 | 0.01 | -4.75    | 0.11  | M1.5 | 0.4 |
| RIK-100 | J16095287-2441535 | 16 09 52.88 | -24 41 53.6 | 0.51 | 0.06 | -15.07   | 0.44  | M5.5 | 0.9 |
| RIK-101 | J16100394-2728479 | 16 10 03.94 | -27 28 48.0 | 0.64 | 0.05 | -7.64    | 0.19  | M4.5 | 0.8 |
| RIK-102 | J16100501-2132318 | 16 10 05.02 | -21 32 31.9 | 0.43 | 0.02 | -54.01   | 1.71  | M0.0 | 0.3 |
| RIK-103 | J16101445-1951377 | 16 10 14.46 | -19 51 37.5 | 0.63 | 0.02 | -4.90    | 0.04  | M4.0 | 0.6 |
| RIK-104 | J16102625-2209105 | 16 10 26.25 | -22 09 10.5 | 0.60 | 0.03 | -7.54    | 0.05  | M3.5 | 0.7 |
| RIK-105 | J16103978-2037094 | 16 10 39.79 | -20 37 09.5 | 0.63 | 0.02 | -4.87    | 0.35  | M2.5 | 0.9 |
| RIK-106 | J16110479-2333166 | 16 11 04.80 | -23 33 16.6 | 0.39 | 0.04 | -1.88    | 0.13  | M0.0 | 0.0 |
| RIK-107 | J16110567-2144032 | 16 11 05.67 | -21 44 03.3 | 0.63 | 0.02 | -6.01    | 0.04  | M1.5 | 0.9 |
| RIK-108 | J16111395-2019188 | 16 11 13.95 | -20 19 18.9 | 0.65 | 0.04 | -12.08   | 0.36  | M3.5 | 0.8 |
| RIK-109 | J16111687-2639331 | 16 11 16.87 | -26 39 33.1 | 0.60 | 0.05 | -5.85    | 0.20  | M5.0 | 0.5 |
| RIK-110 | J16111744-2441203 | 16 11 17.45 | -24 41 20.4 | 0.63 | 0.02 | -5.77    | 0.04  | M4.0 | 0.4 |
| RIK-111 | J16112601-2631558 | 16 11 26.03 | -26 31 55.9 | 0.54 | 0.05 | -3.59    | 0.10  | M2.0 | 0.7 |
| RIK-112 | J16120505-2043404 | 16 12 05.05 | -20 43 40.5 | 0.57 | 0.03 | -12.34   | 0.20  | M1.0 | 1.2 |
| RIK-113 | J16120668-3010270 | 16 12 06.68 | -30 10 27.1 | 0.55 | 0.03 | -33.77   | 0.14  | M0.5 | 0.2 |
| RIK-114 | J16120814-2547578 | 16 12 08.14 | -25 47 57.9 | 0.66 | 0.03 | -6.59    | 0.07  | M3.5 | 0.6 |

| | | | | | | | | | |
|---|---|---|---|---|---|---|---|---|---|
| RIK-115 | J16121368-2431369 | 16 12 13.69 | -24 31 36.9 | 0.31 | 0.03 | -4.25 | 0.05 | M2.5 | 0.6 |
| RIK-116 | J16121723-2839082 | 16 12 17.24 | -28 39 08.6 | 0.61 | 0.06 | -7.14 | 0.58 | M4.5 | 0.2 |
| RIK-117 | J16123352-2543281 | 16 12 33.54 | -25 43 27.6 | 0.69 | 0.06 | -8.52 | 0.22 | M4.0 | 0.6 |
| RIK-118 | J16123531-2034339 | 16 12 35.32 | -20 34 34.0 | 0.46 | 0.03 | -2.30 | 0.07 | M0.5 | 0.8 |
| RIK-119 | J16123604-2723031 | 16 12 36.05 | -27 23 03.2 | 0.39 | 0.03 | 1.06 | 0.03 | K4.0 | 1.2 |
| RIK-120 | J16130627-2606107 | 16 13 06.28 | -26 06 10.7 | 0.38 | 0.03 | -6.11 | 0.07 | M3.5 | 0.7 |
| RIK-121 | J16130978-2044591 | 16 13 09.79 | -20 44 59.1 | 0.57 | 0.03 | -6.11 | 0.04 | M3.5 | 0.9 |
| RIK-122 | J16131008-2435248 | 16 13 10.09 | -24 35 24.8 | 0.65 | 0.06 | -7.67 | 0.33 | M4.5 | 1.0 |
| RIK-123 | J16132054-2229159 | 16 13 20.55 | -22 29 15.9 | 0.38 | 0.02 | -4.18 | 0.07 | M1.5 | 0.4 |
| RIK-124 | J16132190-2136136 | 16 13 21.91 | -21 36 13.7 | 0.53 | 0.03 | -21.99 | 0.20 | M1.5 | 0.2 |
| RIK-125 | J16133279-2044413 | 16 13 32.79 | -20 44 41.4 | 0.48 | 0.02 | -1.96 | 0.03 | M2.5 | 0.8 |
| RIK-126 | J16133644-2326270 | 16 13 36.44 | -23 26 27.0 | 0.36 | 0.02 | -5.87 | 0.04 | M2.5 | 0.4 |
| RIK-127 | J16133834-2158518 | 16 13 38.35 | -21 58 52.0 | 0.43 | 0.04 | -4.62 | 0.12 | M4.5 | 0.8 |
| RIK-128 | J16133840-2443309 | 16 13 38.40 | -24 43 31.2 | 0.33 | 0.02 | -10.56 | 0.07 | M3.5 | 0.6 |
| RIK-129 | J16134490-2434143 | 16 13 44.90 | -24 34 14.4 | 0.47 | 0.02 | -7.23 | 0.05 | M2.5 | 0.7 |
| RIK-130 | J16135663-2457566 | 16 13 56.62 | -24 57 56.8 | 0.36 | 0.02 | -5.08 | 0.04 | M3.5 | 0.7 |
| RIK-131 | J16140380-2453088 | 16 14 03.80 | -24 53 08.9 | 0.42 | 0.02 | -2.48 | 0.06 | M1.0 | 0.6 |
| RIK-132 | J16140733-2217321 | 16 14 07.35 | -22 17 32.2 | 0.57 | 0.05 | -4.97 | 0.18 | M1.0 | 0.0 |
| RIK-133 | J16141011-2217236 | 16 14 10.12 | -22 17 23.5 | 0.62 | 0.04 | -3.09 | 0.04 | M4.0 | 0.7 |
| RIK-134 | J16141259-2454287 | 16 14 12.58 | -24 54 28.8 | 0.49 | 0.02 | -2.06 | 0.03 | M0.5 | 0.6 |
| RIK-135 | J16143363-2004299 | 16 14 33.64 | -20 04 29.9 | 0.55 | 0.02 | -4.44 | 0.08 | M3.5 | 0.6 |
| RIK-136 | J16144125-2556052 | 16 14 41.26 | -25 56 05.3 | 0.67 | 0.02 | -3.99 | 0.05 | M1.5 | 0.7 |
| RIK-137 | J16144989-2139321 | 16 14 49.89 | -21 39 32.2 | 0.56 | 0.02 | -2.61 | 0.03 | M2.5 | 0.4 |
| RIK-138 | J16145244-2513523 | 16 14 52.41 | -25 13 52.4 | 0.16 | 0.03 | -6.54 | 0.50 | M3.5 | 0.6 |
| RIK-139 | J16145269-2308025 | 16 14 52.70 | -23 08 02.7 | 0.42 | 0.03 | -4.23 | 0.04 | M3.5 | 0.6 |
| RIK-140 | J16150060-2919348 | 16 15 00.60 | -29 19 34.9 | 0.28 | 0.03 | -3.54 | 0.09 | M3.0 | 0.2 |
| RIK-141 | J16150620-2500459 | 16 15 06.20 | -25 00 46.1 | 0.56 | 0.05 | -8.56 | 0.47 | M4.5 | 1.3 |
| RIK-142 | J16151104-2322425 | 16 15 11.05 | -23 22 42.7 | 0.56 | 0.03 | -3.39 | 0.03 | M1.0 | 0.4 |
| RIK-143 | J16151239-2318453 | 16 15 12.40 | -23 18 45.3 | 0.30 | 0.05 | -1.15 | 0.11 | K6.0 | 0.0 |
| RIK-144 | J16151948-2540119 | 16 15 19.49 | -25 40 12.0 | 0.66 | 0.04 | -7.58 | 0.14 | M1.5 | 0.7 |
| RIK-145 | J16152750-2627281 | 16 15 27.51 | -26 27 28.1 | 0.46 | 0.04 | -4.09 | 0.06 | M4.0 | 1.0 |
| RIK-146 | J16153220-2010236 | 16 15 32.20 | -20 10 23.7 | 0.60 | 0.02 | -11.06 | 0.15 | M1.5 | 1.0 |
| RIK-147 | J16153311-2707587 | 16 15 33.11 | -27 07 58.8 | 0.58 | 0.03 | -2.55 | 0.06 | M0.5 | 0.6 |
| RIK-148 | J16153587-2529008 | 16 15 35.86 | -25 29 01.0 | 0.41 | 0.02 | -0.13 | 0.02 | K4.5 | 0.7 |
| RIK-149 | J16153642-2622090 | 16 15 36.43 | -26 22 09.2 | 0.36 | 0.03 | -4.29 | 0.08 | M3.5 | 0.0 |
| RIK-150 | J16154732-1911184 | 16 15 47.33 | -19 11 18.5 | 0.57 | 0.02 | -5.08 | 0.12 | M1.5 | 1.2 |
| RIK-151 | J16155987-2325043 | 16 15 59.86 | -23 25 04.5 | 0.49 | 0.03 | -0.80 | 0.04 | M0.0 | 0.0 |
| RIK-152 | J16160080-2214192 | 16 16 00.81 | -22 14 19.3 | 0.51 | 0.04 | -9.47 | 0.33 | M5.5 | 0.1 |
| RIK-153 | J16160292-2430548 | 16 16 02.92 | -24 30 54.8 | 0.47 | 0.05 | -1.92 | 0.07 | M1.0 | 0.8 |
| RIK-154 | J16160856-2041514 | 16 16 08.55 | -20 41 51.5 | 0.56 | 0.03 | -3.81 | 0.10 | M4.0 | 0.6 |
| RIK-155 | J16160947-2243437 | 16 16 09.48 | -22 43 43.8 | 0.39 | 0.02 | -4.15 | 0.04 | M1.5 | 0.5 |
| RIK-156 | J16161720-2609101 | 16 16 17.20 | -26 09 10.2 | 0.71 | 0.05 | -2.24 | 0.06 | K4.0 | 0.3 |
| RIK-157 | J16161893-2542287 | 16 16 18.94 | -25 42 28.7 | 0.61 | 0.04 | -7.34 | 0.14 | M1.5 | 0.7 |
| RIK-158 | J16163346-2552367 | 16 16 33.45 | -25 52 36.8 | 0.53 | 0.02 | -2.84 | 0.04 | M0.5 | 1.3 |
| RIK-159 | J16165130-2433277 | 16 16 51.30 | -24 33 27.7 | 0.14 | 0.02 | 3.59 | 0.06 | F8.0 | 1.4 |
| RIK-160 | J16165984-2154272 | 16 16 59.84 | -21 54 27.3 | 0.17 | 0.02 | -4.22 | 0.05 | M3.5 | 0.4 |
| RIK-161 | J16170606-2225414 | 16 17 06.08 | -22 25 41.2 | 0.31 | 0.04 | -4.78 | 0.18 | M4.5 | 0.9 |
| RIK-162 | J16171380-2251584 | 16 17 13.80 | -22 51 58.5 | 0.58 | 0.04 | -7.15 | 0.20 | M4.5 | 0.4 |
| RIK-163 | J16171649-2327570 | 16 17 16.50 | -23 27 57.1 | 0.61 | 0.02 | -8.59 | 0.05 | M3.5 | 0.6 |
| RIK-164 | J16172162-2325004 | 16 17 21.63 | -23 25 00.4 | 0.64 | 0.03 | -2.00 | 0.07 | M4.0 | 0.5 |
| RIK-165 | J16172297-2121119 | 16 17 22.98 | -21 21 11.9 | 0.60 | 0.02 | -1.45 | 0.05 | M1.0 | 0.0 |
| RIK-166 | J16172615-2450592 | 16 17 26.15 | -24 50 59.2 | 0.45 | 0.02 | -5.71 | 0.04 | M2.5 | 1.3 |
| RIK-167 | J16172769-2421025 | 16 17 27.70 | -24 21 02.6 | 0.55 | 0.04 | -7.42 | 0.66 | M4.5 | 0.6 |
| RIK-168 | J16172994-2451030 | 16 17 29.95 | -24 51 03.2 | 0.34 | 0.02 | -6.97 | 0.04 | M2.5 | 1.4 |
| RIK-169 | J16173031-2438390 | 16 17 30.32 | -24 38 39.2 | 0.58 | 0.03 | -1.72 | 0.05 | M0.0 | 1.6 |
| RIK-170 | J16173138-2303360 | 16 17 31.38 | -23 03 36.0 | 0.24 | 0.02 | 1.05 | 0.03 | K1.5 | 0.0 |
| RIK-171 | J16175501-2711303 | 16 17 55.02 | -27 11 30.4 | 0.46 | 0.02 | -2.95 | 0.04 | M1.5 | 0.2 |
| RIK-172 | J16181997-2005348 | 16 18 19.98 | -20 05 34.9 | 0.33 | 0.03 | 1.24 | 0.04 | K1.5 | 0.4 |
| RIK-173 | J16190214-2138097 | 16 19 02.15 | -21 38 09.7 | 0.43 | 0.02 | -6.13 | 0.05 | M4.5 | 0.2 |
| RIK-174 | J16191473-2532314 | 16 19 14.74 | -25 32 31.3 | 0.34 | 0.02 | 0.59 | 0.02 | K4.5 | 0.7 |

| | | | | | | | | | |
|---|---|---|---|---|---|---|---|---|---|
| RIK-175 | J16191608-2915126 | 16 19 16.09 | -29 15 12.7 | 0.52 | 0.02 | -4.94 | 0.04 | M4.0 | 0.2 |
| RIK-176 | J16193139-2518127 | 16 19 31.39 | -25 18 12.8 | 0.40 | 0.03 | 0.45 | 0.04 | K6.0 | 0.9 |
| RIK-177 | J16194309-2216175 | 16 19 43.10 | -22 16 17.6 | 0.69 | 0.05 | -2.78 | 0.10 | M4.5 | 0.5 |
| RIK-178 | J16194538-2147577 | 16 19 45.38 | -21 47 57.8 | 0.35 | 0.03 | -2.56 | 0.05 | M1.5 | 0.0 |
| RIK-179 | J16194711-2203112 | 16 19 47.12 | -22 03 11.2 | 0.49 | 0.04 | -8.67 | 0.07 | M5.0 | 0.5 |
| RIK-180 | J16194836-2212519 | 16 19 48.36 | -22 12 51.8 | 0.43 | 0.03 | -16.57 | 0.14 | M3.0 | 0.3 |
| RIK-181 | J16194880-2224472 | 16 19 48.78 | -22 24 47.5 | 0.88 | 0.06 | -15.55 | 0.10 | M6.0 | 0.3 |
| RIK-182 | J16194885-2140360 | 16 19 48.86 | -21 40 36.0 | 0.57 | 0.04 | -4.93 | 0.13 | M4.5 | 0.1 |
| RIK-183 | J16200616-2212385 | 16 20 06.16 | -22 12 38.5 | 0.60 | 0.03 | -2.49 | 0.06 | M3.5 | 0.2 |
| RIK-184 | J16200686-2247320 | 16 20 06.86 | -22 47 32.1 | 0.52 | 0.04 | -8.63 | 0.21 | M4.5 | 0.9 |
| RIK-185 | J16202163-2005348 | 16 20 21.64 | -20 05 34.8 | 1.07 | 0.25 | -5.85 | 0.96 | M7.5 | 2.8 |
| RIK-186 | J16202498-2150240 | 16 20 24.98 | -21 50 23.9 | 0.47 | 0.02 | -4.61 | 0.09 | M2.0 | 0.2 |
| RIK-187 | J16202724-2126068 | 16 20 27.24 | -21 26 06.9 | 0.52 | 0.04 | -2.51 | 0.05 | M1.5 | 0.0 |
| RIK-188 | J16203246-2257452 | 16 20 32.46 | -22 57 45.4 | 0.51 | 0.01 | -7.12 | 0.04 | M2.5 | 1.4 |
| RIK-189 | J16203640-2123120 | 16 20 36.41 | -21 23 12.0 | 0.65 | 0.04 | -13.01 | 0.09 | M4.5 | 0.5 |
| RIK-190 | J16204578-2849201 | 16 20 45.79 | -28 49 20.1 | 0.66 | 0.07 | -5.67 | 0.11 | M2.0 | 0.3 |
| RIK-191 | J16205095-2253399 | 16 20 50.95 | -22 53 40.1 | 0.49 | 0.02 | -1.52 | 0.04 | M1.0 | 1.5 |
| RIK-192 | J16211584-2240045 | 16 21 15.84 | -22 40 04.5 | 0.53 | 0.02 | -5.35 | 0.06 | M1.5 | 0.9 |
| RIK-193 | J16212953-2529431 | 16 21 29.55 | -25 29 43.1 | 0.29 | 0.02 | -4.29 | 0.06 | M2.5 | 0.7 |
| RIK-194 | J16212961-2129038 | 16 21 29.62 | -21 29 03.8 | 0.93 | 0.09 | -13.95 | 0.47 | M5.5 | 0.8 |
| RIK-195 | J16214126-2212056 | 16 21 41.26 | -22 12 05.5 | 0.54 | 0.04 | -7.42 | 0.15 | M4.5 | 0.5 |
| RIK-196 | J16214853-2517266 | 16 21 48.54 | -25 17 26.6 | 0.48 | 0.01 | -1.21 | 0.03 | M0.0 | 1.4 |
| RIK-197 | J16215466-2043091 | 16 21 54.67 | -20 43 09.1 | 0.38 | 0.05 | -0.57 | 0.04 | K7 | 0.0 |
| RIK-198 | J16215594-2705034 | 16 21 55.94 | -27 05 03.4 | 0.61 | 0.04 | -7.69 | 0.07 | M5.0 | 0.3 |
| RIK-199 | J16215975-2706366 | 16 21 59.76 | -27 06 36.6 | 0.51 | 0.06 | -5.34 | 0.18 | M4.5 | 0.3 |
| RIK-200 | J16220587-2121556 | 16 22 05.88 | -21 21 55.6 | 0.30 | 0.01 | -2.57 | 0.02 | M1.5 | 0.6 |
| RIK-201 | J16220658-2127089 | 16 22 06.59 | -21 27 08.9 | 0.53 | 0.02 | -7.31 | 0.05 | M3.5 | 0.5 |
| RIK-202 | J16220858-2915063 | 16 22 08.59 | -29 15 06.4 | 0.61 | 0.03 | -3.40 | 0.08 | M2.5 | 0.2 |
| RIK-203 | J16220894-2140372 | 16 22 08.94 | -21 40 37.2 | 0.39 | 0.03 | -4.48 | 0.06 | M3.5 | 0.5 |
| RIK-204 | J16222092-2747095 | 16 22 20.93 | -27 47 09.6 | 0.58 | 0.02 | -3.88 | 0.04 | M2.5 | 0.2 |
| RIK-205 | J16224408-2142222 | 16 22 44.07 | -21 42 22.2 | 0.53 | 0.04 | -3.36 | 0.05 | M3.5 | 0.6 |
| RIK-206 | J16225479-2138091 | 16 22 54.78 | -21 38 09.3 | 0.31 | 0.01 | -5.52 | 0.04 | M2.5 | 0.7 |
| RIK-207 | J16230474-2759252 | 16 23 04.74 | -27 59 25.4 | 0.37 | 0.01 | -1.83 | 0.02 | M0.0 | 0.6 |
| RIK-208 | J16231741-2159067 | 16 23 17.41 | -21 59 07.0 | 0.61 | 0.03 | -5.03 | 0.35 | M3.5 | 0.6 |
| RIK-209 | J16232293-2622161 | 16 23 22.93 | -26 22 16.1 | 0.15 | 0.04 | 1.92 | 0.11 | F5.0 | 0.0 |
| RIK-210 | J16232454-1717270 | 16 23 24.54 | -17 17 27.1 | 0.59 | 0.02 | -5.29 | 0.09 | M2.5 | 0.3 |
| RIK-211 | J16233234-2523485 | 16 23 32.34 | -25 23 48.5 | 0.31 | 0.01 | 0.22 | 0.01 | K2.0 | 1.3 |
| RIK-212 | J16235509-2330396 | 16 23 55.10 | -23 30 40.0 | 0.53 | 0.02 | -3.48 | 0.08 | M2.5 | 1.2 |
| RIK-213 | J16235723-2620244 | 16 23 57.23 | -26 20 24.6 | 0.33 | 0.02 | -7.36 | 0.05 | M4.5 | 0.3 |
| RIK-214 | J16235790-2602296 | 16 23 57.91 | -26 02 29.9 | 0.39 | 0.02 | -1.01 | 0.03 | M0.0 | 0.9 |
| RIK-215 | J16235902-2736037 | 16 23 59.02 | -27 36 03.7 | 0.47 | 0.02 | -6.44 | 0.04 | M3.5 | 0.2 |
| RIK-216 | J16240942-2134075 | 16 24 09.43 | -21 34 07.3 | 0.33 | 0.11 | -5.41 | 0.30 | M3.5 | 1.1 |
| RIK-217 | J16241177-1955583 | 16 24 11.77 | -19 55 58.3 | 0.51 | 0.01 | -4.78 | 0.04 | M1.5 | 1.2 |
| RIK-218 | J16241551-2544344 | 16 24 15.52 | -25 44 34.5 | 0.31 | 0.02 | -0.04 | 0.02 | M0.0 | 1.2 |
| RIK-219 | J16241860-2854475 | 16 24 18.60 | -28 54 47.5 | 0.37 | 0.02 | 0.85 | 0.03 | G8.5 | 1.3 |
| RIK-220 | J16250236-2321447 | 16 25 02.37 | -23 21 45.1 | 0.65 | 0.03 | -3.08 | 0.05 | M1.0 | 1.8 |
| RIK-221 | J16251690-2322030 | 16 25 16.90 | -23 22 03.6 | 0.63 | 0.03 | -3.62 | 0.09 | M0.0 | 1.9 |
| RIK-222 | J16252327-2727314 | 16 25 23.27 | -27 27 31.5 | 0.49 | 0.03 | -8.88 | 0.06 | M5.0 | 0.2 |
| RIK-223 | J16252883-2607538 | 16 25 28.81 | -26 07 53.8 | 0.55 | 0.02 | -2.86 | 0.04 | M2.5 | 0.9 |
| RIK-224 | J16253274-2611386 | 16 25 32.74 | -26 11 38.6 | 0.72 | 0.06 | -8.74 | 0.34 | M5.5 | 0.8 |
| RIK-225 | J16253504-2332549 | 16 25 35.04 | -23 32 55.0 | 0.58 | 0.03 | -2.18 | 0.07 | M0.0 | 2.1 |
| RIK-226 | J16254808-2154195 | 16 25 48.09 | -21 54 19.5 | 0.42 | 0.06 | -4.53 | 0.42 | M3.5 | 0.2 |
| RIK-227 | J16254925-2554371 | 16 25 49.24 | -25 54 37.3 | 0.47 | 0.02 | -1.40 | 0.03 | M0.0 | 1.2 |
| RIK-228 | J16255541-2721242 | 16 25 55.41 | -27 21 24.3 | 0.73 | 0.04 | -8.91 | 0.23 | M5.0 | 0.1 |
| RIK-229 | J16255790-2600374 | 16 25 57.91 | -26 00 37.3 | 0.31 | 0.06 | -0.14 | 0.08 | G8.0 | 1.3 |
| RIK-230 | J16261964-2137207 | 16 26 19.64 | -21 37 20.8 | 0.43 | 0.02 | -1.07 | 0.03 | M1.5 | 0.8 |
| RIK-231 | J16261998-2233023 | 16 26 19.98 | -22 33 02.5 | 0.48 | 0.03 | -1.24 | 0.04 | M1.5 | 0.8 |
| RIK-232 | J16261995-2258097 | 16 26 19.98 | -22 58 10.0 | 0.60 | 0.03 | -3.06 | 0.07 | M3.5 | 0.7 |
| RIK-233 | J16262016-2233124 | 16 26 20.16 | -22 33 12.7 | 0.34 | 0.04 | -4.53 | 0.09 | M3.5 | 0.8 |
| RIK-234 | J16262803-2526477 | 16 26 28.04 | -25 26 47.8 | 0.52 | 0.02 | -2.00 | 0.03 | M0.0 | 1.8 |

| | | | | | | | | | |
|---|---|---|---|---|---|---|---|---|---|
| RIK-235 | J16263276-2622589 | 16 26 32.77 | -26 22 59.0 | 0.58 | 0.12 | -5.88 | 0.53 | M6.0 | 1.2 |
| RIK-236 | J16263495-2511409 | 16 26 34.95 | -25 11 40.9 | 0.48 | 0.01 | -1.01 | 0.02 | M0.0 | 1.1 |
| RIK-237 | J16264120-2200094 | 16 26 41.21 | -22 00 09.5 | 0.35 | 0.02 | -2.80 | 0.08 | M1.5 | 0.7 |
| RIK-238 | J16270950-2618549 | 16 27 09.51 | -26 18 54.9 | 0.48 | 0.04 | -8.58 | 0.10 | M4.5 | 0.5 |
| RIK-239 | J16271273-2504017 | 16 27 12.74 | -25 04 01.8 | 0.53 | 0.02 | -9.14 | 0.12 | M1.0 | 0.9 |
| RIK-240 | J16272766-2821502 | 16 27 27.66 | -28 21 50.4 | 0.56 | 0.03 | -5.38 | 0.15 | M2.0 | 0.3 |
| RIK-241 | J16273320-2821097 | 16 27 33.22 | -28 21 09.8 | 0.13 | 0.01 | -4.93 | 0.05 | M3.5 | 0.2 |
| RIK-242 | J16275794-2524187 | 16 27 57.93 | -25 24 18.8 | 0.32 | 0.03 | -6.06 | 0.13 | M3.5 | 0.1 |
| RIK-243 | J16284604-2711574 | 16 28 46.05 | -27 11 57.4 | 0.33 | 0.02 | -2.52 | 0.08 | M1.5 | 1.1 |
| RIK-244 | J16295662-2659182 | 16 29 56.63 | -26 59 18.4 | 0.52 | 0.03 | -8.00 | 0.08 | M4.5 | 0.1 |
| RIK-245 | J16300275-2727006 | 16 30 02.76 | -27 27 00.5 | 0.42 | 0.04 | -3.65 | 0.19 | M1.5 | 0.1 |
| RIK-246 | J16310579-2725460 | 16 31 05.80 | -27 25 46.1 | 0.38 | 0.04 | -9.05 | 0.11 | M4.5 | 0.3 |
| RIK-247 | J16311542-2657151 | 16 31 15.42 | -26 57 15.1 | 0.45 | 0.02 | -2.12 | 0.06 | M1.5 | 0.7 |
| RIK-248 | J16315668-2846126 | 16 31 56.69 | -28 46 12.7 | 0.17 | 0.04 | -9.99 | 0.10 | M4.0 | 0.7 |
| RIK-249 | J16323587-1612577 | 16 32 35.87 | -16 12 57.8 | 0.13 | 0.01 | 0.90 | 0.02 | K4.5 | 2.3 |
| RIK-250 | J16333496-1832540 | 16 33 34.97 | -18 32 53.9 | 0.17 | 0.05 | -8.47 | 0.21 | M3.5 | 0.4 |
| RIK-251 | J16333503-2715448 | 16 33 35.04 | -27 15 44.9 | 0.35 | 0.03 | -3.86 | 0.25 | M3.5 | 0.4 |
| RIK-252 | J16333881-2150263 | 16 33 38.83 | -21 50 26.3 | 0.73 | 0.02 | -1.70 | 0.03 | M1.5 | 1.3 |
| RIK-253 | J16343514-2658030 | 16 34 35.14 | -26 58 03.1 | 0.54 | 0.05 | -27.10 | 0.70 | M4.5 | 0.5 |
| RIK-254 | J16343826-2835504 | 16 34 38.27 | -28 35 50.5 | 0.22 | 0.02 | -3.31 | 0.04 | M1.0 | 0.8 |
| RIK-255 | J16350625-2025282 | 16 35 06.26 | -20 25 28.3 | 0.61 | 0.04 | -11.45 | 0.14 | M4.5 | 0.6 |
| RIK-256 | J16354573-2711166 | 16 35 45.74 | -27 11 16.6 | 0.26 | 0.02 | -8.03 | 2.47 | M1.5 | 0.0 |
| RIK-257[a] | J16134781-2747340 | 16 13 47.82 | -27 47 34.0 | 0.09 | 0.02 | 4.14 | 0.04 | F4.5 | 0.2 |



\end{document}